\documentclass[11pt]{article}
\usepackage{latexsym} \usepackage{epsf}
\usepackage{a4}
\usepackage{amsfonts}
\usepackage{amsmath}
\usepackage{amsthm}
\usepackage{amssymb}
\usepackage{graphics}
\usepackage{cite}
\renewcommand{\theequation}{\thesection.\arabic{equation}}
\newcounter{subequation}[equation]
\makeatletter

\expandafter\let\expandafter\reset@font\csname reset@font\endcsname

\def\subeqnarray{\arraycolsep1pt
    \def\@eqnnum\stepcounter##1{\stepcounter{subequation}%
        {\reset@font\rm(\theequation\alph{subequation})}}
\jot5mm     \eqnarray}

\makeatother

\def\tr{\mathop{\hbox{\rm tr}}\nolimits}

\def\be{\begin{equation}}
\def\ee{\end{equation}}
\def\bea{\begin{eqnarray}}
\def\eea{\end{eqnarray}}
\def\dd{\partial}
\def\half{\frac{1}{2}}

\def\one#1{#1^{\raise5pt\hbox{$\scriptstyle\!\!\!\!1$}}\,{}}
\def\two#1{#1^{\raise5pt\hbox{$\scriptstyle\!\!\!\!2$}}\,{}}

\def\II{\hbox{{1}\kern-.25em\hbox{l}}}
\makeatletter
\def\binrel@#1{\begingroup
  \setboxz@h{\thinmuskip0mu
    \medmuskip\m@ne mu\thickmuskip\@ne mu
    \setbox\tw@\hbox{$#1\m@th$}\kern-\wd\tw@
    ${}#1{}\m@th$}%
  \edef\@tempa{\endgroup\let\noexpand\binrel@@
    \ifdim\wdz@<\z@ \mathbin
    \else\ifdim\wdz@>\z@ \mathrel
    \else \relax\fi\fi}%
  \@tempa
}
\let\binrel@@\relax
\def\overset#1#2{\binrel@{#2}%
  \binrel@@{\mathop{\kern\z@#2}\limits^{#1}}}
\def\underset#1#2{\binrel@{#2}%
  \binrel@@{\mathop{\kern\z@#2}\limits_{#1}}}
\makeatother
\newfont{\bbd}{msbm10 scaled\magstep1}
\def\C{\hbox{\bbd C}}

\def\NN{\hbox{\bbd N}}
\def\R{\hbox{\bbd R}}
\def\V{\hbox{\bbd V}}

\def\P{\hbox{\bbd P}}

\begin{document}
\begin{titlepage}

\vspace*{1cm}

\begin{center}
{\LARGE \bf{ Baxter operators with deformed symmetry }}

\vspace{1cm}

{\large \sf D. Chicherin$^{ea}$\footnote{\sc e-mail:chicherin@pdmi.ras.ru},
  S. Derkachov$^{a}$\footnote{\sc e-mail:derkach@pdmi.ras.ru}, D.
Karakhanyan$^{bc}$\footnote{\sc e-mail:karakhan@lx2.yerphi.am},
R. Kirschner$^d$\footnote{\sc e-mail:Roland.Kirschner@itp.uni-leipzig.de} \\
}

\vspace{0.5cm}

\begin{itemize}
\item[$^a$]
{\it St. Petersburg Department of Steklov Mathematical Institute
of Russian Academy of Sciences,
Fontanka 27, 191023 St. Petersburg, Russia}
\item[$^b$]
{\it Yerevan Physics Institute, \\
Br. Alikhanian st. 2,  Yerevan, 0036, Armenia}
\item[$^c$]
{\it Yerevan State University, 1 Alex Manoogian St., Yerevan, 0025,
Armenia}
\item[$^d$]
{\it Institut f\"ur Theoretische
Physik, Universit\"at Leipzig, \\
PF 100 920, D-04009 Leipzig, Germany}
\item[$^e$]
{\it Chebyshev Laboratory, St.-Petersburg State University,\\
14th Line, 29b, Saint-Petersburg, 199178 Russia}
\end{itemize}
%\vspace{4cm}
%{\bf Abstract}
\end{center}
\vspace{0.5cm}
\begin{abstract}
Baxter operators are constructed for quantum spin chains with deformed
$s\ell_2$ symmetry. The parallel treatment of Yang-Baxter operators for the
cases of undeformed, trigonometrically and elliptically deformed symmetries
presented earlier and relying on the factorization regarding parameter
permutations is extended to the global chain operators following the
scheme  worked out recently in the undeformed case.
\end{abstract}

\vspace{4cm}

\end{titlepage}

{\small \tableofcontents}
\renewcommand{\refname}{References.}
\renewcommand{\thefootnote}{\arabic{footnote}}
\setcounter{footnote}{0} \setcounter{equation}{0}

%\begin{center} \LARGE
%Some calculations on the general $\mathrm{R}$-operator for
%$U_q(s\ell_2)$-symmetric spin chain
%\end{center}

%%%%%%%%%%%%%%%%%%%%%%%%%%%%%%%%%%%%%%%%%%%%%%%%%%%%%%%%%%%%%%%%%%%%%%%%%%%%%%%%%%%%%%%%%%%%
%%%%%%%%%%%%%%%%%%%%%%%%%%%%%%%%%%%%%%%%%%%%%%%%%%%%%%%%%%%%%%%%%%%%%%%%%%%%%%%%%%%%%%%%%%%%
%%%%%%%%%%%%%%%%%%%%%%%%%%%%%%%%%%%%%%%%%%%%%%%%%%%%%%%%%%%%%%%%%%%%%%%%%%%%%%%%%%%%%%%%%%%%
\section{Introduction}
\setcounter{equation}{0}
%%%%%%%%%%%%%%%%%%%%%%%%%%%%%%%%%%%%%%%%%%%%%%%%%%%%%%%%%%%%%%%%%%%%%%%%%%%%%%%%%%%%%%%%%%%%

We consider periodic quantum spin chains with integrable dynamics where the
single-site quantum states form an infinite-dimensional irreducible
representation of the trigonometrically or elliptically deformed $s\ell_2$
symmetry algebra characterized by the representation parameter or spin
$\ell$. The set of commuting quantum observables can be represented by the
transfer matrix $\mathrm{t}(u)$ in a generating function form. The task of
finding the spectrum and eigenstates of these conserved charges can be
treated by the  algebraic Bethe ansatz method. The Bethe equation is related
to the Baxter equation, being a difference equation involving the transfer
matrix and the Baxter operator $\mathrm{Q}(u)$.
%%%%%%%%---------------------------------------------------------
The concept of Baxter operators has been introduced by Baxter \cite{Baxter}
in analyzing the eight-vertex model.
Baxter operators have been constructed for various models, e.g. in
\cite{BzSt90,GP92,Volkov,Vol1,BLZ,SDQ,KSS,Pronko,Zab,Backlund,RW,KMS}.
General algebraic schemes of construction have been formulated in
\cite{Feigin,BLZ}. The case of non-compact representations has been
addressed in particular in \cite{Volkov,Vol1,BLZ,SDQ,KiMa,Kor,ByTe,DerMa}.
This concept provides an alternative
way of solution, in particular a Baxter operator allows to construct the
separated variable representation \cite{Sklyanin,DKMI}.
%%-----------------------------------------------------------------

In a number of papers a systematization and reformulation of known results
on integrable quantum systems and essential progress has been achieved in view
of the case of non-compact representations appearing in application to gauge
field theory.  In \cite{CDKKI} the approach of constructing Baxter operators
based on general Yang-Baxter operators has been presented in detail for the
undeformed $s\ell_2$ symmetry. The
non-compact representation case is understood as the generic one and the
case of finite-dimensional representation is obtained in the limit where
$2\ell$ approaches non-negative integer values.
The relation of this to another approach worked out in \cite{BLMS} has been
investigated in detail \cite{CDKKII}.

%% remove this%%
%For reviews of the basic literature and some basic relations we refer to these
%papers \cite{BS,CDKKI}.

The general Yang-Baxter operators serve as local building blocks for global
spin chain operators.
Their construction has been formulated in \cite{ParamPerm}
in a uniform way for the three cases of the symmetry algebra $s\ell_2$
undeformed, trigonometrically and elliptically deformed. The factorization
regarding the permutation of representation parameters is the essential
feature of this construction.
The features of factorization have been noticed earlier in studies of the
chiral Potts model \cite{Kashaev,Bazh,Miwa,Tarasov}.
By the results of \cite{EllDerSp} the formulation of the elliptic case has been
essentially completed. The  particular permutation
operator intertwining representation $\ell$ and $-\ell-1$
of Sklyanin algebra was taken
in \cite{ParamPerm} in the  form of a series relying on \cite{Zab}
(see also \cite{Zab11}) which is
well defined rather in the finite-dimensional case but not in the
infinite-dimensional one. The form based on
the elliptic beta integral \cite{Sp,spi:essays,Spiridonov10} 
avoids this problem and is easier to handle.

In the present paper we extend the parallel treatment of the three cases
from the operators related to the chain sites to the global chain operators.
For the results in \cite{CDKKI} referring to the generic infinite-dimensional
representation case we present here the corresponding extensions to these
two cases of deformation.
The constructions in \cite{CDKKI} result in  explicit expressions for the
relevant operators and their action. Analogous explicit results are given
here for the deformed cases.

In Section 2 a summary of the relevant relations of \cite{ParamPerm,CDKKI} is given
formulating simultaneously the relations which hold uniformly in all three
cases.
We present the general scheme which is suited for the three cases of
symmetry algebra and allows to construct Baxter $\mathrm{Q}$-operators,
{\it general} transfer matrices
and to establish their {\it commutativity} and {\it factorization} properties.
As a new result we derive a formula relating
both Baxter $\mathrm{Q}$-operators.
In the Sections \ref{q} and \ref{ellip} we specify the
general formulae and work out the details for the cases of $q$-deformation and
elliptic deformation, respectively.
Some more details, less important for comparison of the three cases, but
useful and potentially  important in further investigations, have been put into the
Appendix.

\section{Factorized Yang-Baxter operators and Baxter equation}
\setcounter{equation}{0}
%%%%%%%%%%%%%%%%%%%%%%%%%%%%%%%%%%%%%%%%%%%%%%%%%%%%%%%%%%%%%%%%%%%%%%%%%%%%%%%%%%%%%%%%%%%%%%%%%%%%%%%%%%%
Building blocks in the construction of the quantum systems of
integrable spin chains are the $\mathrm{R}$-operators
depending on spectral parameter $u$,
intertwining the tensor product of two representations
\be \label{RR}
\mathbb{R}_{12}(u):
\mathbb{V}_{1}\otimes\mathbb{V}_{2} \to \mathbb{V}_{1}\otimes\mathbb{V}_{2}
\ee
and obeying the Yang-Baxter
relation,
\begin{equation} \label{YB}
\R_{12}(u-v) \, \R_{13}(u) \,
\R_{23}(v) =
\R_{23}(v) \, \R_{13}(u) \,
\R_{12}(u-v)\,.
\end{equation}
All operators act in the tensor product of three spaces
$\V_{1}\otimes\V_{2}\otimes\V_{3}$, where e.g.  $\R_{12}$ acts
non-trivially in the tensor product of the first and the second spaces (\ref{RR}) and as
identity operator on the remaining space $\V_{3}$.

We consider irreducible representations of $s\ell_2$ and its trigonometric and
elliptic deformations which are parameterized by spin $\ell$.
The representations are infinite dimensional for generic complex number
$\ell$, i.e. for $\ell \neq \frac{n}{2},\,n=0,1,2\cdots$, but finite $n+1$ dimensional for
$\ell = \frac{n}{2}$. Initially (\ref{YB}) is written without restrictions on
the representations involved. In the following we shall preserve the notation
$\R_{12}(u)$ for the
{\it general} $\mathrm{R}$-operator which by definition acts on the
tensor product of two infinite dimensional representations
$$
\mathbb{R}_{12}(u|\ell_1,\ell_2):
\mathbb{V}_{\ell_1}\otimes\mathbb{V}_{\ell_2} \to \mathbb{V}_{\ell_1}\otimes\mathbb{V}_{\ell_2}
$$
and respects the Yang-Baxter relation (\ref{YB})
in the space $\mathbb{V}_{\ell_1}\otimes\mathbb{V}_{\ell_2}\otimes\mathbb{V}_{\ell_3}$
\begin{equation} \label{YBE}
\R_{12}(u-v|\ell_1,\ell_2) \, \R_{13}(u|\ell_1,\ell_3) \,
\R_{23}(v|\ell_2,\ell_3) =
\R_{23}(v|\ell_2,\ell_3) \, \R_{13}(u|\ell_1,\ell_3) \,
\R_{12}(u-v|\ell_1,\ell_2)\,.
\end{equation}
Along with the previous case the cases of $\mathrm{R}$-operators with one or
both tensor factors finite dimensional are to be considered. If one factor
is the fundamental spin $\half$ representation $\C^2$ the $\mathrm{R}$-operator is called
$\mathrm{L}$-operator
and if both are the fundamental spin $\half$ representation then one has the
fundamental $\mathcal{R}$-matrix,
$
\mathcal{R}(u): \C^2\otimes\C^2 \to \C^2\otimes\C^2
$. The latter can be considered as the source
of the relevant algebra and co-algebra relations. Indeed, the particular
case of the relation (\ref{YB}) where  the representations labeled by
$1,2$ are the fundamental ones $\mathbb{V}_1=\mathbb{V}_2=\C^2$,
\begin{equation} \label{Yangian}
\mathcal{R}_{ij,nm}(u-v) \, \mathrm{L}_{ns}(u) \, \mathrm{L}_{mp}(v) =
\mathrm{L}_{is}(v) \,\mathrm{L}_{jp}(u) \, \mathcal{R}_{sp,nm}(u-v)
\end{equation}
%%%%%%%%%%%%%%%%%%%%%%%%%%%%%%%%%%%%%%%%%%%%%%%%%%%%%%%%%%%%%%%%%%%%%%%%%%%%%%%%%%%%%
%%\begin{equation}\label{fundRLL}
%%\mathcal{R}(u-v)\,\mathrm{L}(u)\otimes \mathrm{L}(v)=
%%\mathrm{L}(v)\otimes \mathrm{L}(u)\,\mathcal{R}(u-v)\,.
%%\end{equation}
%%%%%%%%%%%%%%%%%%%%%%%%%%%%%%%%%%%%%%%%%%%%%%%%%%%%%%%%%%%%%%%%%%%%%%%%%%%%%%%%%%%%%%%
the given $\mathcal{R}$ fixes the commutation relations of the matrix
elements of the involved $\mathrm{L}$-operator.
%%Here $\otimes$ denotes the matrix tensor product.
Here summation over indices $i,j,\cdots = 1,2$ is assumed.
We have in mind the situation where
these matrix elements generate the generic irreducible representation related
to one site of the spin chain.
$$
\mathrm{L}(u|\ell) : \mathbb{V}_{\ell}\otimes\C^2 \to \mathbb{V}_{\ell}\otimes\C^2
$$
It is distinguished from other solutions by a rather simple dependence on
spectral parameter $u$ and  linearity in the generators
of the symmetry algebra.

The co-algebra structure is also defined because the relation
(\ref{Yangian}) still holds if a solution $\mathrm{L}(u)$ is substituted by
the matrix product
$\mathrm{L}_1(u)\,\mathrm{L}_2(u) \cdots\mathrm{L}_N(u)$ acting on the tensor product
representation related to the sites $1,2, \cdots, N$. Besides of the case
related to irreducible representations of $s\ell_2$ or its trigonometric and elliptic
deformations
other solutions of (\ref{Yangian}) with the same $\mathcal{R}$  are known,
which emerge as degeneracy limits of the former. They have been considered
recently in \cite{BLMS} and also in \cite{CDKKII}
concerning the undeformed $s\ell_2$; the present study will not touch this case.

Further, the case of (\ref{YB})
if representations labeled by $1,2$ are generic spin $\ell_1$
and spin $\ell_2$ representations correspondingly $\V_{1}=\V_{\ell_1}$,
$\V_2 =\V_{\ell_2}$ but the
representation $3$ is fundamental $\V_3=\C^2$
\begin{equation}\label{RRLL}
\R_{12}(u-v|\ell_1,\ell_2)\,\mathrm{L}_1(u|\ell_1)\, \mathrm{L}_2(v|\ell_2)=
\mathrm{L}_2(v|\ell_2)\, \mathrm{L}_1(u|\ell_1)\,\R_{12}(u-v|\ell_1,\ell_2)\,
\end{equation}
can be read as the defining relation for the general $\mathrm{R}$-operator
with the given $\mathrm{L}$-operator.

At this point it is convenient to adopt  notations
showing  that representations of the symmetry algebra
with parameters $\ell$ and $-\ell-1$ are equivalent, since the corresponding
values of Casimir operators are equal.  For this reason we join
the spectral parameter $u$ and spin parameter $\ell$ into two independent
linear combinations $u_1,u_2$ such that
%% \footnote{For example in the case of
%% undeformed $s\ell_2$ considered in \cite{CDKKI} $u_1=u-\ell-1$, $u_2 = u+\ell$.}
 \begin{equation} \label{ul}
u_1 \leftrightarrow u_2  \;\;\;\sim\;\;\; \ell \leftrightarrow -\ell-1
\end{equation}
and  denote the $\mathrm{L}$-operator also by $\mathrm{L}(u_1,u_2)$.
For example in the case of
undeformed $s\ell_2$ considered in \cite{CDKKI} we have $u_1=u-\ell-1$, $u_2 = u+\ell$.
In the cases of deformed symmetries the corresponding relations will be
specified below.
Further  we refer to both $u_1,u_2$ as spectral parameters.
Correspondingly
the general $\mathrm{R}$-operator in (\ref{RRLL})
appears as depending on the parameters $u_1,u_2, v_1,v_2$ where
$$
u_1 \leftrightarrow u_2  \;\;\;\sim\;\;\; \ell_1 \leftrightarrow -\ell_1-1
\;\;\;\;;\;\;\;\;
v_1 \leftrightarrow v_2  \;\;\;\sim\;\;\; \ell_2 \leftrightarrow -\ell_2-1\,.
$$
Rewriting now (\ref{RLL}) in terms of $\mathrm{R}_{12} = \mathrm{P}_{12} \R_{12}$,
where $\mathrm{P}_{12}$ is the operator of permutation of the
tensor factors,
\begin{equation} \label{RLL}
\mathrm{R}_{12}(u_1,u_2|v_1,v_2) \, \mathrm{L}_1(u_1,u_2) \, \mathrm{L}_2(v_1,v_2) =
\mathrm{L}_1(v_1,v_2) \, \mathrm{L}_2(u_1,u_2) \, \mathrm{R}_{12}(u_1,u_2|v_1,v_2)
\end{equation}
we see the action of $\mathrm{R}$-operator on the product of
$\mathrm{L}$-operators appearing as the permutation a pair
of parameters $(u_1,u_2)$ in the first space $\mathrm{L}$-operator
with a pair $(v_1,v_2)$ in the second space $\mathrm{L}$-operator
and represents a permutation $s$ in the set of four parameters
$$
s: \mathbf{u} = ({v_1,v_2},{u_1,u_2}) \mapsto
({u_1,u_2},{v_1,v_2})\,.
$$
It turns out to be useful to study operators corresponding to other
permutation operations on this set of parameters.
An arbitrary permutation of four parameters can be constructed out of
three elementary transpositions $s^i$ (i=1,2,3)
interchanging  a pair of adjacent parameters only.
For them we find an operator representation $s^i \mapsto \mathrm{S}^i(\mathbf{u})$
with the composition rule
$s^i s^j \mapsto \mathrm{S}^i(s^j\mathbf{u})\,\mathrm{S}^j(\mathbf{u})$
such that
$$
\mathrm{S}^i(\mathbf{u}) \, \mathrm{L}_1(u_1,u_2) \, \mathrm{L}_2(v_1,v_2) =
\mathrm{L}_1(u'_1,u'_2) \, \mathrm{L}_2(v'_1,v'_2) \, \mathrm{S}^i(\mathbf{u}) \;\;;\;\;
s^i \mathbf{u} = (u'_1,u'_2,v'_1,v'_2)
$$
Furthermore, operator relations corresponding to the
symmetric group defining relations
$s^i s^i = \II$, $s^1 s^2 s^1 = s^2 s^1 s^2$, $s^3 s^2 s^3 = s^2 s^3 s^2$
have to be satisfied.
The operators $\mathrm{S}^1,\,\mathrm{S}^2,\,\mathrm{S}^3$ have been constructed
for the three cases of
symmetry algebra in \cite{ParamPerm,EllDerSp}.
Each operator effectively depends only on one parameter.
Their defining relations have the form
\begin{eqnarray} \label{S1}
\mathrm{S}^1(v_2-v_1) \, \mathrm{L}_2 (v_1,v_2) &=& \mathrm{L}_2 (v_2,v_1) \, \mathrm{S}^1(v_2-v_1)\,, \\[0.2 cm]
                 \label{S2LL}
\mathrm{S}^2(u_1-v_2) \; \mathrm{L}_1 (u_1 , u_2) \; \mathrm{L}_2 (v_1 , v_2) &=&
\mathrm{L}_1 (v_2 , u_2) \; \mathrm{L}_2 (v_1 , u_1) \; \mathrm{S}^2(u_1-v_2)\,, \\ [0.2 cm]
                 \label{S3}
\mathrm{S}^3(u_2-u_1) \, \mathrm{L}_1 (u_1,u_2) &=& \mathrm{L}_1 (u_2,u_1) \, \mathrm{S}^3(u_2-u_1)\,.
\end{eqnarray}
The binary relation $\mathrm{S}^i(a) \,\mathrm{S}^i (-a) = \II$ and
the triple Coxeter relations
\begin{eqnarray}
                \label{S1S2S1}
\mathrm{S}^1(a) \, \mathrm{S}^2(a+b) \, \mathrm{S}^1(b) =
\mathrm{S}^2(b) \, \mathrm{S}^1(a+b) \, \mathrm{S}^2(a)\,, \\ [0.2 cm]
                \label{S3S2S3}
\mathrm{S}^3(a) \, \mathrm{S}^2(a+b) \, \mathrm{S}^3(b) =
\mathrm{S}^2(b) \, \mathrm{S}^3(a+b) \, \mathrm{S}^2(a)
\end{eqnarray}
do hold implying that we have indeed an
operator representation of the symmetric group.
Particulary $\mathrm{S}^i(0)=\II$.
Since the defining relations (\ref{S1}) and (\ref{S3}) are essentially identical
and in view of (\ref{ul}) the operators $\mathrm{S}^1(a)$ and $\mathrm{S}^3(a)$ are two
copies of the intertwining operator $\mathrm{W}(a)$ of the symmetry algebra,
\begin{equation} \label{WS}
\mathrm{W}(u_2-u_1) \, \mathbf{S}_a(\ell) =
\mathbf{S}_a(-\ell-1)\,\mathrm{W}(u_2-u_1),
\end{equation}
where $\mathbf{S}_a(\ell)$ denote
spin $\ell$ representation of the generators of the symmetry algebra,
acting in the second $\mathrm{S}^1(a) = \mathrm{W}_2(a)$ and
the first $\mathrm{S}^3(a) = \mathrm{W}_1(a)$ quantum spaces.
It is worth mentioning that at (half)-integer $\ell$ ($u_2 -u_1=
2\ell +1 \in \NN$)
the intertwining operator $\mathrm{W}$ has a non-trivial kernel
coinciding with the $2\ell +1$ dimensional invariant subspace.
This will be shown below by rewriting this operator in a particular form
valid in this case.
 For consistency of (\ref{WS})  the vectors
annihilated by $\mathrm{W}$ have to span  an invariant subspace.

Out of elementary transposition operators more involved
operators can be constructed. We need the operators $\mathrm{R}^1$ and $\mathrm{R}^2$
which satisfy
the defining relations
%% \footnote{Here we underline the parameters to be transposed
%% for readers convenience.}
\begin{eqnarray}
                  \label{R1LL}
\mathrm{R}^{1}(u_1|v_1,v_2) \, \mathrm{L}_1(u_1,u_2) \,
                               \mathrm{L}_2(v_1,v_2) &=&
\mathrm{L}_1(v_1,u_2) \, \mathrm{L}_2(u_1,v_2) \,
\mathrm{R}^{1}(u_1|v_1,v_2)\,, \\ [0.2 cm]
                  \label{R2LL}
\mathrm{R}^{2}(u_1,u_2|v_2) \, \mathrm{L}_1(u_1,u_2) \,
                               \mathrm{L}_2(v_1,v_2) &=&
\mathrm{L}_1(u_1,v_2) \, \mathrm{L}_2(v_1,u_2) \,
\mathrm{R}^{2}(u_1,u_2|v_2)
\end{eqnarray}
and can be factorized
%%, for example,
as follows
\begin{eqnarray}
                            \label{R1}
\mathrm{R}^{1}(u_1|v_1,v_2) = \mathrm{S}^2 (v_2-v_1) \, \mathrm{S}^1 (u_1-v_1) \, \mathrm{S}^2 (u_1-v_2)
                            = \mathrm{S}^1 (u_1-v_2) \, \mathrm{S}^2 (u_1-v_1) \, \mathrm{S}^1 (v_2-v_1)\,, \\ [0.2 cm]
                            \label{R2}
\mathrm{R}^{2}(u_1,u_2|v_2) = \mathrm{S}^2 (u_2-u_1) \, \mathrm{S}^3 (u_2-v_2) \, \mathrm{S}^2 (u_1-v_2)
                            = \mathrm{S}^3 (u_1-v_2) \, \mathrm{S}^2 (u_2-v_2) \, \mathrm{S}^3 (u_2-u_1)\,.
\end{eqnarray}
Finally using $\mathrm{R}^1$ and $\mathrm{R}^2$ we factorize the general $\mathrm{R}$-operator
\begin{equation} \label{R}
\mathrm{R}(u_1,u_2|v_1,v_2) = \mathrm{R}^1(u_1|v_1,u_2) \, \mathrm{R}^2(u_1,u_2|v_2) =
\mathrm{R}^2(v_1,u_2|v_2) \, \mathrm{R}^1(u_1|v_1,v_2)\,.
\end{equation}
Notice that we have two factorized
representations for the $\mathrm{R}$-operator and that
%% and $\mathrm{R}^1$, $\mathrm{R}^2$.
their consistency follows from Coxeter relations (\ref{S1S2S1}) and (\ref{S3S2S3}).

Now we proceed to global operators building them out of local ones
considered above. It is well known that the physics of a homogeneous periodic spin
chains can
be obtained from the transfer matrix defined from the $\mathrm{L}$-operators by
\begin{equation}
\label{tmat} \mathrm{t}(u) = \tr
\mathrm{L}_1(u)\,\mathrm{L}_2(u)\cdots \mathrm{L}_N(u)\,,
\end{equation}
where the lower
index $k$ refers to the  local quantum
space at the $k$-th site and the trace is taken over an auxiliary space $\C^2$,
 because it is the generating function of the set of commuting operators,
$[\,\mathrm{t}(u)\,,\, \mathrm{t}(v)\,] = 0$, as the consequence of (\ref{Yangian}).
In the homogeneous case we restrict to $\ell_1 = \ell_2 = ...= \ell_N = \ell$.

In a similar manner we build the {\it general} transfer matrix
substituting  locally the  $\mathrm{L}$-operators by general $\mathrm{R}$-operators
%%and taking the trace over infinite dimensional auxiliary space $\mathbb{V}_s$
\begin{equation} \label{T}
\mathrm{T}_{s}(u|\ell) = \tr_{0} \R_{10} (u|\ell, s)\,\R_{20} (u|\ell,
s) \cdots \R_{N0} (u|\ell, s)\,.
\end{equation}
Whereas the matrix trace in (\ref{tmat}) concerns the
fundamental representation now the trace is to be taken in the generic
infinite dimensional representation $\mathbb{V}_s$ of spin $s$
labeled by index 0.
We assume $\ell$ to be fixed and often omit it using
the  notation $\mathrm{T}_{s}(u)$.
Similarly out of the local operators $\mathrm{R}^1$ and $\mathrm{R}^2$
we build the following traces of monodromies
\begin{eqnarray}\label{Q1}
\mathrm{Q}_1(u-v_1|\ell) = \tr_{0}\, \R^{1}_{10}(u_{1}|v_1,u_{2})\cdots
\R^{1}_{N0}(u_{1}|v_1,u_{2})\,, \\ [0.2 cm]
\label{Q2}
\mathrm{Q}_2(u-v_2|\ell) = \tr_{0}\, \R^{2}_{10}(u_{1},u_2|v_{2})\cdots
\R^{2}_{N0}(u_{1},u_2|v_{2})\,.
\end{eqnarray}
The form of the  spectral parameter dependence in the two previous formulae
follows from corresponding property of local building blocks
and can be seen from (\ref{R1}) and (\ref{R2}).

The introduced operators happen to be {\it commutative}
\begin{equation} \label{commut}
[\,\mathrm{t}(u) , \mathrm{Q}_i(v)\, ] = 0 \ \ ; \ \
[ \, \mathrm{T}_{s}(u) , \mathrm{Q}_k (v) \, ] = 0 \ \ ;\ \ [ \,
\mathrm{Q}_i (u) , \mathrm{Q}_k (v) \, ] = 0 \ \ ;\ \ [ \,
\mathrm{P} , \mathrm{Q}_{k}(u) \, ] = 0 \ \ ;\ \ i,k = 1,2\,
\end{equation}
where $\mathrm{P} =\mathrm{P}_{12}\mathrm{P}_{13}\cdots\mathrm{P}_{1N}$ is the cyclic
permutation along the closed chain, and they respect {\it factorization} relations
\begin{equation}\label{Factor}
\mathrm{P}\cdot\mathrm{T}_{s} (u-v) =
\mathrm{Q}_1(u-v_1)\,\mathrm{Q}_2(u-v_2) =
\mathrm{Q}_2(u-v_2)\,\mathrm{Q}_1(u-v_1)
\end{equation}
where $v_1,v_2$ are linear combinations of $v$ and $s$ analogous
to the case of $u_1,u_2,u,\ell$ (\ref{ul}).
Corresponding proofs rely on
Yang-Baxter like relations and can be found in \cite{CDKKI}.

In \cite{ParamPerm} we have obtained uniformly the general Yang-Baxter operators
and their factors related to parameter permutations in the cases of
$s\ell_2$ symmetry undeformed and with quantum and elliptic deformation.
Starting from the well-known fundamental $\mathcal{R}$-matrices for each case,
we have formulated the $\mathrm{L}$-operators with the matrix elements embedded in the
algebra generated by Heisenberg conjugated pairs $z, \dd $. We have noticed the
factorized form in all considered cases
\be \label{Lfact}
\mathrm{L}(u_1,u_2) = [u]\, \mathrm{V}^{-1} (z,u_2)\, \mathrm{D}(z,\dd)\, \mathrm{V}(z,u_1)\,
,
\ee
where $[u]$ is a function of spectral parameter, $\mathrm{V}$ and $\mathrm{D}$ are
two by two matrices with operator entries.
The formulation of the spin chain  takes  such  pairs $z_i, \dd_i, i=1, ..., N $
for each site and an additional one for the auxiliary space labeled as $i=
0$. Our operator constructions rely first on the  algebraic relations, based on the
algebra generated by the mentioned canonical pairs. The results appear as
expressions in these generators or as related integral operators with
integration over $z_i$.
We did not investigate all aspects of detailed
definition of these constructions  as operators in functional spaces with bilinear
forms, i.e.  using the word operator in view of the related physics
we are not claiming a construction completed in the latter sense.

As suggested by the notation we consider representations by functions of
$z$. In the undeformed and $q$-deformed cases the  representation modules
of spin $\ell$ can be considered as spanned by monomials of non-negative
powers with 1 as lowest weight vector. Whereas the above algebraic
construction is done in the frame of Heisenberg pairs in the elliptic
case as well we do not provide a detailed description of the representations
in this case as embedded in  Heisenberg algebra representations.
The known facts about these representation are mentioned in Section 4.
Actually, the detailed form of the representation is not needed in the construction.
The above trace should be defined by the embedding in the Heisenberg algebra
representation.

In the next step we have to prove that the introduced traces of monodromies
$\mathrm{Q}_1$ and $\mathrm{Q}_2$ (\ref{Q1}), (\ref{Q2}) are indeed Baxter
$\mathrm{Q}$-operators. For   this goal it remains to check
that they do respect Baxter equation. Above
we have derived factorization and commutativity properties
of operators $\mathrm{T}_s$, $\mathrm{Q}_1$, $\mathrm{Q}_2$ from
appropriate local relations for their building blocks.
Similarly  we derive the Baxter equation from a local relation.
Starting from the defining relation for $\mathrm{R^2}$ (\ref{R2LL})
we shall obtain the following local relation in the space
$\mathbb{V}_{\ell}\otimes\mathbb{V}_{s}\otimes\C^2$
\begin{equation} \label{cornerstone}
\mathrm{Z}^{-1}_{0} \cdot
\mathbb{R}^2_{k0}(u) \,\mathrm{L}_k (u_1 ,u_2)\cdot
\mathrm{Z}_{0}
=
\begin{pmatrix}
\kappa^{-1} \cdot \mathbb{R}^2_{k0}(u + \delta) & \cdots \\
0 & \kappa \, \Delta(u_1,u_2)\cdot \mathbb{R}^2_{k0}(u - \delta)
\end{pmatrix}\,.
\end{equation}
Here we use the shorthand notation $\mathbb{R}^2(u) =
\mathbb{R}^2(u_1,u_2|0)$. The
index $k$ refers to the corresponding local quantum space $\mathbb{V}_{\ell}$
in the spin chain site,
and the index $0$ refers to the infinite dimensional auxiliary space
$\mathbb{V}_{s}$.
$\mathrm{Z}_0$ denotes a certain auxiliary matrix acting
in the space $\mathbb{V}_{s}\otimes\C^2$. $\kappa$ and $\delta$ are some constants
and $\Delta(u_1,u_2)$ is a {\it symmetric} function of the spectral parameters:
$\Delta(u_1,u_2)=\Delta(u_2,u_1)$. In the underformed case
we have $\Delta(u_1,u_2) = (u_1 u_2)^N$, where $N$ is the number of sites,
and in the deformed cases the corresponding deformed modifications of this expression appears.
The matrix element above the diagonal denoted by ellipsis in (\ref{cornerstone})
is not indicated explicitly since we do not need it for our purposes.
Detailed calculations leading to (\ref{cornerstone}) will be done in
Subsections \ref{qBaxEq} and \ref{ellipBaxEq} for the cases of $q$-deformation
and elliptic deformation, respectively.
Considerations in both cases
%%are based on (\ref{R2LL}) and
follow the general strategy and are very similar as in \cite{CDKKI}.
They  do not use the explicit expression for operator $\mathrm{R}^2$
or for its building blocks $\mathrm{S}^2$, $\mathrm{S}^3$, but only
their properties. The calculations in both cases  use only:
\begin{enumerate}
\item The defining relation for $\mathrm{R}^2$ (\ref{R2LL}).
\item The second factorization of $\mathrm{R}^2$ (\ref{R2}) in the product of elementary permutation operators.
%%\begin{equation} \label{R22}
%%? why repetition ?
%%\mathrm{R}^2(u_1 , u_2 | v_2) = \mathrm{S}^3 (u_1 - v_2) \, \mathrm{S}^2 (u_2 - v_2) \, \mathrm{S}^3 (u_2 - u_1)
%%\end{equation}
\item The factorization formula for $\mathrm{L}$-operator (\ref{Lfact}).
\item The general property of the
operator $\mathrm{R}^2$: $[\,\mathrm{R}^2_{12} ,z_2\,] =0$ that
follows from (\ref{R2}).
\item Several {\it recurrence relations} for %%builduing blocks of $\mathrm{R}^2$ --
$\mathrm{S}^2$ and $\mathrm{S}^3$, connecting $\mathrm{S}^i(a\pm\delta)$ with
$\mathrm{S}^i(a)$ at $i=2,3$.
\end{enumerate}

As a by-product we obtain a set of peculiar recurrence relations
for intertwining operators of the symmetry algebra.
These relations are similar to the recurrence relations 
used for the evaluation of the q-beta-integral\cite{aar} and elliptic beta-integral \cite{Sp,spi:essays}.

In the elliptic case the recurrence relations give rise to a factorized form
for the intertwining operator suitable for
finite-dimensional representations
of the Sklyanin algebra. It is an alternative form
to the one proposed by A.~Zabrodin in \cite{Zab98}.

Having the local relation (\ref{cornerstone}) it is rather straightforward
to produce the corresponding global relation.
We form the monodromy
$\mathbb{R}^2_{10}(u) \cdots \mathbb{R}^2_{N0}(u)\, \mathrm{L}_1 (u) \cdots \mathrm{L}_N (u)$,
apply $N$ times the local relation (\ref{cornerstone}) obtaining the product of $N$ triangular matrices
with operator entries and calculate the traces over the auxiliary two
dimensional space $\C^2$
and the auxiliary infinite dimensional space $\mathbb{V}_{s}$ and obtain
the Baxter equation for $\mathrm{Q}_2(u)$
\begin{equation} \label{BE}
\mathrm{t}(u)\, \mathrm{Q}_2(u) = \kappa^{-N}\, \mathrm{Q}_2(u+\delta) +
\kappa^{N}\, \Delta^N(u_1,u_2)\,\mathrm{Q}_2(u-\delta)\,.
\end{equation}

Similarly starting from the local relation for $\mathrm{R}^1$ (\ref{R1LL}) it
is possible
to obtain the Baxter equation for $\mathrm{Q}_1(u)$.
However, going beyond the review of \cite{CDKKI},
here we  establish a useful relation between
the two  Baxter operators $\mathrm{Q}_1$ and $\mathrm{Q}_2$.
To make the presentation more transparent we use the following notations
for the elementary transposition operators
\begin{equation} \label{S1S2S3}
\mathrm{S}^1(a) = \mathrm{W}_2(a) \;\;;\;\;
\mathrm{S}^2(a) = \mathrm{S}_{12}(a)\;\;;\;\;
\mathrm{S}^3(a) = \mathrm{W}_1(a)\,,
\end{equation}
where we take into account (\ref{WS}). Lower indices
on the right hand side  refer to spaces
where the corresponding operators act nontrivially.
Taking into account (\ref{R2})
we rewrite (\ref{Q2}) as follows
\begin{eqnarray}
\nonumber
\mathrm{Q}_2(u|\ell) &=& \tr_{0}\, \P_{10}\, \mathrm{W}_1 (u_1)\, \mathrm{S}_{10}(u_2)\, \mathrm{W}_1 (u_2-u_1) \cdots
\P_{N0} \, \mathrm{W}_N (u_1) \, \mathrm{S}_{N0}(u_2) \, \mathrm{W}_N (u_2-u_1) = \\[0.2 cm]
\label{Q2Tau}
&=& \tr_{0}\, \P_{10}\, \mathrm{W}_1 (u_1)\, \mathrm{S}_{10}(u_2) \cdots
\P_{N0} \, \mathrm{W}_N (u_1) \, \mathrm{S}_{N0}(u_2) \cdot
\mathcal{T}(u_2-u_1)\,,
\end{eqnarray}
where we have introduced the operator
$$
\mathcal{T}(a) = \mathrm{W}_1 (a) \cdots \mathrm{W}_N (a)
$$
which is a product of intertwining operators referring to the
$N$ sites of the spin chain.
Consequently  the property $\mathcal{T}(a) \mathcal{T}(-a) = \II$ holds.
Similarly due to (\ref{R1}) the first $\mathrm{Q}$-operator (\ref{Q1}) takes the form
\begin{eqnarray}
\nonumber
\mathrm{Q}_1(u|{\scriptstyle -\ell-1}) &=&
\tr_{0}\, \P_{10}\, \mathrm{W}_0 (u_2-u_1)\, \mathrm{S}_{10}(u_2)\, \mathrm{W}_0 (u_1) \cdots
\P_{N0} \, \mathrm{W}_0 (u_2-u_1) \, \mathrm{S}_{N0}(u_2) \, \mathrm{W}_0 (u_1) = \\[0.2 cm]
\label{Q1Tau}
&=& \mathcal{T}( u_2-u_1) \cdot \tr_{0}\, \P_{10} \,\mathrm{S}_{10}(u_2)\, \mathrm{W}_0 (u_1) \cdots
\P_{N0} \,\mathrm{S}_{N0}(u_2)\, \mathrm{W}_0 (u_1)\,.
\end{eqnarray}
Notice that in $\mathrm{Q}_1$ we choose the representation parameter in quantum space
to be $-\ell-1$
but not $\ell$ that corresponds to the transposition of spectral parameters
$u_1 \leftrightarrow u_2$ (\ref{ul}). Then taking into
account the cyclicity of the trace it is easy to see that the traces of monodromies
in (\ref{Q2Tau}) and (\ref{Q1Tau}) coincide.
Thus we conclude that the two Baxter operators are related
by the similarity transformation
\begin{equation} \label{QTau}
\mathcal{T}(u_2-u_1)\,\mathrm{Q}_2(u|\ell) =
\mathrm{Q}_1(u|{\scriptstyle -\ell-1})\,\mathcal{T}(u_2-u_1)
\end{equation}
and consequently the Baxter equation for $\mathrm{Q}_1(u)$
has exactly the same form as (\ref{BE}).

Finally in \cite{CDKKI} we have shown that the trace of
any monodromy of the form
\begin{equation}
\label{A} \mathbf{A} =
\mathrm{P}_{10}\,\mathrm{A}_{}(z_1,\partial_1|z_0)\cdot\mathrm{P}_{20}\,
\mathrm{A}_{}(z_2,\partial_2|z_0)\, \cdots\,
\mathrm{P}_{N0}\,\mathrm{A}_{}(z_N,\partial_N|z_0)
\end{equation}
can be easily calculated
\begin{equation} \label{trA}
\tr_{_0}\mathbf{A} = \mathrm{P}\cdot\left.\mathrm{A}_{}(z_1,\partial_1|z_2)\cdot
\mathrm{A}_{}(z_2,\partial_2|z_3)\cdots
\mathrm{A}_{}(z_N,\partial_N|z_0)\right|_{z_0\to z_1}\,.
\end{equation}
%%%%%%%%%%%%%%%%%%%%%%%%%%%%%%%%%%%%%%%%%%%%%%%%%%%%%%%%%%%%%%%%%%%%%%%%%%%%%%%%%%%%%%%%%%%
The same formula is valid if $\mathbf{A}$ is an integral operator.
The Baxter operator $\mathrm{Q}_2$ constructed out of $\mathrm{R}^2$
fits into this formula since, as we have already mentioned,
$\mathrm{R}^2_{12}$ commutes with the variable $z_2$.

In analogy to the undeformed case,
using (\ref{trA}) in $q$-case and elliptic case we shall obtain an explicit formula for
the action of $\mathrm{Q}_2(u)$ on a particular state appearing as a generating function
by its dependence on  auxiliary parameters.

\section{Trigonometric deformation case} \label{q}
%%$\mathcal{U}_q(s\ell_2)$ symmetry algebra
\setcounter{equation}{0}
%%%%%%%%%%%%%%%%%%%%%%%%%%%%%%%%%%%%%%%%%%%%%%%%%%%%%%%%%%%%%%%%%%%%%%%%%%%%%%%%%%%%%%%%%%%%%%%%%%%%%%%%%%%%%%%%%%%
%%%%%%%%%%%%%%%%%%%%%%%%%%%%%%%%%%%%%%%%%%%%%%%%%%%%%%%%%%%%%%%%%%%%%%%%%%%%%%%%%%%%%%%%%%%%%%%%%%%%%%%%%%%%%%%%%%%
\subsection{Trigonometric $\mathrm{L}$-operator and parameter permutation operators}
%%%%%%%%%%%%%%%%%%%%%%%%%%%%%%%%%%%%%%%%%%%%%%%%%%%%%%%%%%%%%%%%%%%%%%%%%%%%%%%%%%%%%%%%%%%%%%%%%%%%%%%%%%%%%%%%%%%

Now we are going to
apply the above strategy  to the $q$-deformed symmetry  $\mathcal{U}_q(s\ell_2)$.
We are interested in infinite dimensional
representations of the algebra
on Verma modules, i.e. the representation space $\mathbb{V}_{\ell}$
coincides with the space of polynomials $\C[z]$ and the generators of the algebra are
realized as finite-difference operators. Details can be found in \cite{ParamPerm}.
%%Substituting operator in (\ref{L})
%%it is not difficult to see \cite{ParamPerm} that
%%$\mathrm{L}$-operator takes
%%factorized form
The $\mathrm{L}$-operator respecting (\ref{Yangian})
has the factorized form
\begin{equation}
\label{Lq}
\mathrm{L}(u_1,u_2) =
\begin{pmatrix}
1 & 1 \\
z q^{-u_2} & z q^{u_2}
\end{pmatrix}
\begin{pmatrix}
q^{z\partial_z+1} & 0 \\
0 & q^{-z \partial_z-1}
\end{pmatrix}
\begin{pmatrix}
q^{u_1} & -z^{-1} \\
- q^{-u_1} & z^{-1}
\end{pmatrix}
\end{equation}
where the two sets of parameters are related
according to (\ref{ul}) as
\begin{equation} \label{u}
u_1 =u -\ell -1 \; ; \; u_2 = u+\ell\,.
\end{equation}
Further we quote the
operators of elementary permutations $\mathrm{S}^1$,
$\mathrm{S}^2$, $\mathrm{S}^3$ being the
building blocks of the general $\mathrm{R}$-operator
and constructed in \cite{ParamPerm}.
\begin{itemize}
\item
$\mathrm{S}^1(a)$ and $\mathrm{S}^3(a)$ (\ref{S1}), (\ref{S3})
are two copies of the operator $\mathrm{W}(a)$ acting
nontrivially in the
second and the first quantum spaces, respectively (\ref{S1S2S3}).
It has the explicit form
\begin{equation} \label{qW}
\mathrm{W}(a) = \frac{q^{\frac{a^2}{2}}}{z^a} \cdot
\frac{(q^{2 z \partial_z + 2 - 2 a};q^2)}{(q^{2 z \partial_z + 2};q^2)}
\cdot q^{-a z\partial_z}\,.
\end{equation}
In the latter formula $(x;q^2)$ denotes the  infinite $q$-product (\ref{qprod}).
$\mathrm{W}(2\ell+1)$ intertwines
representations parameterized by $\ell$ and $-\ell -1$ (\ref{WS}).
Let us mention that in addition to the
indispensable symmetric group relations we have
$\mathrm{W}(a)\mathrm{W}(-a) = \II$ and  furthermore the exponential property:
$\mathrm{W}(a)\mathrm{W}(b) = \mathrm{W}(a+b)$.
\item
The operator $\mathrm{S}^2$ defined by (\ref{S2LL}) acts nontrivially
in the tensor products of two quantum spaces.
We choose it in the following form
\begin{equation} \label{qS2}
\mathrm{S}^2(a) = z^a_1 \cdot
\frac{(\frac{z_2}{z_1}q^{1-a};q^2)}{(\frac{z_2}{z_1}q^{1+a};q^2)}\,.
\end{equation}
\end{itemize}
%%%%%%%%%%%%%%%%%%%%%%%%%%%%%%%%%%%%%%%%%%%%%%%%%%%%%%%%%%%%%%%%%%%%%%%%%%%%%%%%%%%%%%%%%%%%%%%%%%%%%%%%%%%%%%%%%%%

Let us stress that the defining relations (\ref{WS}) and (\ref{S2LL}) do not
fix uniquely the operators $\mathrm{W}(a)$ and $\mathrm{S}^2(a)$. Indeed,
 let us consider the equation for the operator $\mathrm{S}(\mathbf{u})$
\begin{equation} \label{intw}
\mathrm{S}(\mathbf{u}) \cdot \mathrm{L}_1(u_1,u_2) \, \mathrm{L}_2(v_1,v_2) =
\mathrm{L}_1(u'_1,u'_2) \, \mathrm{L}_2(v'_1,v'_2) \cdot \mathrm{S}(\mathbf{u})
\end{equation}
where $(v'_1,v'_2,u'_1,u'_2)$ is a permutation
of the set $\mathbf{u}=(v_1,v_2,u_1,u_2)$. It is clear that if we multiply
any solution $\mathrm{S}$
of the latter equation by an arbitrary operator
$\varphi$ such that
\begin{equation} \label{period}
[\,z_1\,,\,\varphi\,] = [\,z_2 \,,\,\varphi\,] = 0 \;\;;\;\;
[\,q^{2 z_1 \partial_{z_1}} \,,\,\varphi\,] = [\,q^{2 z_2 \partial_{z_2}}\, , \,\varphi\,] = 0
\end{equation}
we obtain another solution of (\ref{intw}). In particular we are free to
multiply (\ref{qS2}) by an arbitrary multiplicatively-periodic function
$\varphi(z_1,z_2)$ with
multiplicative period equal to $q^2$:
$\varphi(q^2 z_1,z_2) = \varphi(z_1,q^2 z_2) = \varphi(z_1,z_2)$.
In this way we find another solution of (\ref{S2LL})
which happens to be useful for us,
\begin{equation} \label{qS2'}
\mathrm{S}'^2(a) = z^a_2 \cdot
\frac{(\frac{z_1}{z_2}q^{1-a};q^2)}{(\frac{z_1}{z_2}q^{1+a};q^2)}\,.
\end{equation}

Having on disposal the  operators of elementary permutations
we find the operators $\mathrm{R}^1$, $\mathrm{R}^2$.
In \cite{DKK} they  have been constructed
in another way solving directly
the system of operator relations (\ref{R1LL}), (\ref{R2LL}).
In Appendix \ref{R1R2} we show that the
two constructions do agree.

In \cite{ParamPerm} Coxeter relation (\ref{S1S2S1}), (\ref{S3S2S3})
have been proved using the series expansion for (\ref{qW}) and the
$q$-summation formula.
In Appendix \ref{qCoxeter} we show that Coxeter relations follow  from
the  pentagon formula (\ref{pentagon}) only.

The general $\mathrm{R}$-operator concerns the generic
 symmetry algebra representations $\ell_1, \ell_2$. The analytic dependence
on these representation parameters
 contains in the limits to integer or half-integer values
 simpler objects like $\mathrm{L}$-operator.
%%%%that is implied by the hierarchy (\ref{hierarchy})
In \cite{CDKKI} we have shown how to extract the $\mathrm{L}$-operator
from $\mathbb{R}(u|\ell,s)$ at $s=\frac{1}{2}$ restricting it to
invariant subspace $\mathbb{V}_{\ell}\otimes\C^2$ in the case of undeformed
$s\ell_2$ symmetry.
In Appendix \ref{R->L} we present the  analogous calculation
in the case of $q$-deformation.
%%%%%%%%%%%%%%%%%%%%%%%%%%%%%%%%%%%%%%%%%%%%%%%%%%%%%%%%%%%%%%%%%%%%%%%%%%%%%%%%%%%%%%%%%%%%%%%%%%%%%%%%%
%%%%%%%%%%%%%%%%%%%%%%%%%%%%%%%%%%%%%%%%%%%%%%%%%%%%%%%%%%%%%%%%%%%%%%%%%%%%%%%%%%%%%%%%%%%%%%%%%%%%%%%%%
%%%%%%%%%%%%%%%%%%%%%%%%%%%%%%%%%%%%%%%%%%%%%%%%%%%%%%%%%%%%%%%%%%%%%%%%%%%%%%%%%%%%%%%%%%%%%%%%%%%%%%%%%%
%%%%%%%%%%%%%%%%%%%%%%%%%%%%%%%%%%%%%%%%%%%%%%%%%%%%%%%%%%%%%%%%%%%%%%%%%%%%%%%%%%%%%%%%
%%%%%%%%%%%%%%%%%%%%%%%%%%%%%%%%%%%%%%%%%%%%%%%%%%%%%%%%%%%%%%%%%%%%%%%%%%%%%%%%%%%%%%%%
%%%%%%%%%%%%%%%%%%%%%%%%%%%%%%%%%%%%%%%%%%%%%%%%%%%%%%%%%%%%%%%%%%%%%%%%%%%%%%%%%%%%%%%%
%%%%%%%%%%%%%%%%%%%%%%%%%%%%%%%%%%%%%%%%%%%%%%%%%%%%%%%%%%%%%%%%%%%%%%%%%%%%%%%%%%%%%%%%
%%%%%%%%%%%%%%%%%%%%%%%%%%%%%%%%%%%%%%%%%%%%%%%%%%%%%%%%%%%%%%%%%%%%%%%%%%%%%%%%%%%%%%%%
%%%%%%%%%%%%%%%%%%%%%%%%%%%%%%%%%%%%%%%%%%%%%%%%%%%%%%%%%%%%%%%%%%%%%%%%%%%%%%%%%%%%%%%%
\subsection{Trigonometric recurrence relations}
\label{qRecurr}
%%%%%%%%%%%%%%%%%%%%%%%%%%%%%%%%%%%%%%%%%%%%%%%%%%%%%%%%%%%%%%%%%%%%%%%%%%%%%%%%%
Now we are going to establish several recurrence relations,
which relate the elementary permutation operators with shifted arguments, i.e.
we show how to connect $\mathrm{S}^{i}(a)$ and $\mathrm{S}^{i}(a\pm 1)$ ($i=1,2,3$).
We shall need such relations for proving Baxter equation.

First we consider the recurrence relations for $\mathrm{S}^1$, $\mathrm{S}^3$
which are in fact two copies of the operator $\mathrm{W}(a)$.
We have
\begin{equation}
\label{rec}
- q^{-\frac{1}{2}} \, \mathrm{W} (a + 1) =
            \mathrm{W} (a) \, \frac{1}{z}
            \left( q^{z \partial_{z}} - q^{-z \partial_{z}} \right) =
            \frac{1}{z} \left( q^{z \partial_{z}} - q^{-z \partial_{z}} \right)
\mathrm{W} (a)\,.
\end{equation}
It can be easily checked using the explicit expression (\ref{qW}).

In passing we notice that starting from  $\mathrm{W}(0)=\II$
we obtain  a factorized representation
for $\mathrm{W}(n)$ if $n$ is nonnegative integer
\begin{equation}
\mathrm{W}(n) =
q^{\frac{n}{2}}
\left[
\frac{1}{z}
\left( q^{-z \partial_{z}} - q^{z \partial_{z}} \right)
\right]^n\,.
\end{equation}
By this it is easy to see that the action of $\mathrm{W}(n)$ annihilates
$z^k, k=0,1,...,n-1$ spanning the invariant subspace in the case $2\ell+1 = n
\in \NN$.

Further we quote two matrix relations,
\begin{eqnarray}
\label{rec1}
- q^{\frac{1}{2}} (q^{a} - q^{-a}) \, \mathrm{W}(a-1)
\begin{pmatrix}
1 ,& 1
\end{pmatrix} & = &
\begin{pmatrix}
-z & 1
\end{pmatrix} \mathrm{W}(a)
\begin{pmatrix}
1 & 1 \\
z q^{-a} & z q^{a}
\end{pmatrix}
\begin{pmatrix}
q^{z \partial_{z} + 1} & 0 \\
0 & q^{-z \partial_{z} - 1}
\end{pmatrix},
\\ [0.2 cm]
\label{rec2}
q^{\frac{1}{2}} (q^{a} - q^{-a}) \, z^{-1}\,\mathrm{W}(a-1)
\begin{pmatrix}
1 \\ -1
\end{pmatrix} & = &
\begin{pmatrix}
q^{z \partial_{z} + 1} & 0 \\
0 & q^{-z \partial_{z} - 1}
\end{pmatrix}
\begin{pmatrix}
q^{a} & -z^{-1} \\
-q^{-a} & z^{-1}
\end{pmatrix}
\mathrm{W}(a)
\begin{pmatrix}
1 \\ z
\end{pmatrix}.
\end{eqnarray}
In order to see that the right hand side of (\ref{rec1}) is
proportional to the row
$\begin{pmatrix} 1 , & 1\end{pmatrix}$
we multiply the intertwining relation (\ref{WS})
$
\mathrm{W}(a) \, \mathrm{L}(0,a) =
\mathrm{L}(a,0) \, \mathrm{W}(a)
$
by the row
$\begin{pmatrix}
 -z ,& 1
\end{pmatrix}$ on the left obtaining
$
\begin{pmatrix}
-z ,& 1
\end{pmatrix} \mathrm{W}(a) \, \mathrm{L}(0,a) =
\begin{pmatrix}
0 , & 0
\end{pmatrix}
$
which is equivalent (\ref{Lq}) to
$$
\begin{pmatrix}
-z ,& 1
\end{pmatrix} \mathrm{W}(a)
\begin{pmatrix}
1 & 1 \\
z q^{-a} & z q^{a}
\end{pmatrix}
\begin{pmatrix}
q^{z \partial_{z} + 1} & 0 \\
0 & q^{-z \partial_{z} - 1}
\end{pmatrix}
\begin{pmatrix}
1 \\ -1
\end{pmatrix} = 0\,,
$$
and this implies the linear dependence
of the two equations in the system (\ref{rec1}).
Consequently verifying the system of two relations
in (\ref{rec1})
one needs to check only the first one which can be done easily
taking into account the explicit expression (\ref{qW}) for $\mathrm{W}(a)$.
Similarly  the second system (\ref{rec2}) of operator
relations can be proven. In fact verifying Baxter equation we shall
need  (\ref{rec2}) only.

The needed recurrence relation for $\mathrm{S}^2$ (\ref{qS2}) are
\begin{eqnarray} \label{useful3}
q^{z_2 \partial_2} \, \mathrm{S}^2 (a) \, q^{-z_2 \partial_2} & = &
z_1^{-1} \,\begin{pmatrix} 1 - \frac{z_2}{z_1} \, q^{-a} \end{pmatrix}^{-1} \mathrm{S}^2 (a + 1)\,, \\ [0.2 cm]
                 \label{useful6}
q^{z_2 \partial_2} \, \mathrm{S}^2 (a) \, q^{-z_2 \partial_2} & = & (z_1 - z_2 q^a) \, \mathrm{S}^2 (a - 1)\,, \\ [0.2 cm]
                 \label{useful7}
q^{-z_2 \partial_2} \, \mathrm{S}^2 (a) \, q^{z_2 \partial_2} & = & (z_1 - z_2 q^{-a}) \, \mathrm{S}^2 (a - 1)\,.
\end{eqnarray}
The latter simple relations could  be considered not worth to be displayed
here. However they have direct analogs
in the elliptic case which are much more involved as we shall see in
Subsection \ref{recurrEl}.
%%%%%%%%%%%%%%%%%%%%%%%%%%%%%%%%%%%%%%%%%%%%%%%%%%%%%%%%%%%%%%%%%%%%%%%%%%%%%%%%%%%%%%%%
%%%%%%%%%%%%%%%%%%%%%%%%%%%%%%%%%%%%%%%%%%%%%%%%%%%%%%%%%%%%%%%%%%%%%%%%%%%%%%%%%%%%%%%%
%%%%%%%%%%%%%%%%%%%%%%%%%%%%%%%%%%%%%%%%%%%%%%%%%%%%%%%%%%%%%%%%%%%%%%%%%%%%%%%%%%%%%%%%
\subsection{Trigonometric Baxter equation}
\label{qBaxEq}
%%%%%%%%%%%%%%%%%%%%%%%%%%%%%%%%%%%%%%%%%%%%%%%%%%%%%%%%%%%%%%%%%%%%%%%%%%%%%%%%%
Now we are ready to proceed to the Baxter equation for $\mathrm{Q}_2$ (\ref{Q2}).
It is constructed from several copies of local operator $\mathrm{R}^2$ which
respects the relation (\ref{R2LL}) interchanging parameters
$u_2 \leftrightarrow v_2$ in the product of two $\mathrm{L}$-operators.
Further it will be convenient for us to
exploit its second factorized representation in (\ref{R2}).
%%\begin{equation} \label{R22}
%%\mathrm{R}^2(u_1 , u_2 | v_2) = \mathrm{S}_3 (u_1 - v_2) \, \mathrm{S}_2 (u_2 - v_2) \, \mathrm{S}_3 (u_2 - u_1)
%%\end{equation}

We derive the Baxter equation from the appropriate local relation for
building blocks of the transfer matrix $\mathrm{t}(u)$ (\ref{tmat})
and of $\mathrm{Q}_2$ (\ref{Q2}).
%%and  we calculate the action of
%%operator $\mathrm{R}^2$ \eqref{R22} onto $\mathrm{L}$-operator (\ref{LFact}).
More exactly we find expressions for the
diagonal elements of the matrix (cf. (\ref{cornerstone}))
\begin{equation} \label{qRL}
\begin{pmatrix}
1 & 0 \\
-z_1 & 1
\end{pmatrix}
\mathrm{R}^2_{12}(u_1 , u_2 | v_2 ) \, \mathrm{L}_1 (u_1 , u_2)
\begin{pmatrix}
1 & 0 \\
z_2 & 1
\end{pmatrix}
\end{equation}
in the limit $v_2 = 0$.
As we shall see shortly in the examined limit the matrix element below  diagonal
turns to zero.
%%The operator expression (\ref{eq:RL}) acts in the space $\mathbb{C}^2\otimes V_1 \otimes V_2$
%%where $V_i \approx \mathbb{C}[z_i]$ ($i=1,2$).
%%The operator $\mathrm{R}^2_{12}$ acts nontrivially in the tensor product $V_1 \otimes V_2$
%%of two infinite dimensional spaces, $\mathrm{L}$-operator as well as
%%the leftmost triangular matrix in (\ref{eq:RL}) acts in $\mathbb{C}^2\otimes V_1$ and
%%the rightmost triangular matrix acts in $\mathbb{C}^2\otimes V_2$.
Starting from \eqref{R2LL}, substituting the factorized form of the
$\mathrm{L}$-operator
\eqref{Lq} and taking into account the commutativity of
$\mathrm{R}^2$ with $z_2$ one obtains
\begin{equation} \label{qbasic}
\begin{array}{c}
\mathrm{R}^2(u_1 , u_2 |v_2) \, \mathrm{L}_1(u_1 , u_2)
\begin{pmatrix}
1 & 1 \\ z_2 q^{-v_2} & z_2 q^{v_2}
\end{pmatrix}
=
\mathrm{L}_1 (u_1 , v_2)
\begin{pmatrix}
1 & 1 \\ z_2 q^{-u_2} & z_2 q^{u_2}
\end{pmatrix} \times
\\ [0.5 cm]
\cdot
\begin{pmatrix}
q^{z_2 \partial_2} \, \mathrm{R}^2(u_1 , u_2 |v_2) \, q^{-z_2 \partial_2} & 0 \\
 0 & q^{-z_2 \partial_2} \, \mathrm{R}^2(u_1 , u_2 | v_2) \, q^{z_2 \partial_2}
\end{pmatrix}
\end{array}
\end{equation}
or in the other form
\begin{equation} \label{qbasic2}
\begin{array}{c}
\mathrm{R}^2(u_1 , u_2 |v_2) \, \mathrm{L}_1(u_1 , u_2)
=
\mathrm{L}_1 (u_1 , v_2)
\begin{pmatrix}
1 & 1 \\ z_2 q^{-u_2} & z_2 q^{u_2}
\end{pmatrix} \times
\\ [0.5 cm]
\cdot
\begin{pmatrix}
q^{z_2 \partial_2} \, \mathrm{R}^2(u_1 , u_2 |v_2) \, q^{-z_2 \partial_2} & 0 \\
 0 & q^{-z_2 \partial_2} \, \mathrm{R}^2(u_1 , u_2 | v_2) \, q^{z_2 \partial_2}
\end{pmatrix}
\begin{pmatrix}
z_2 q^{v_2} & -1 \\
-z_2 q^{-v_2} & 1
\end{pmatrix}
\frac{1}{z_2(q^{v_2} - q^{-v_2})}\,.
\end{array}
\end{equation}
These are the two main relations in the present calculation.

We start with the  matrix element below diagonal in (\ref{qRL}).
Taking into account \eqref{Lq} and
$$
\begin{pmatrix}
 -z_1 , & 1
\end{pmatrix}
\mathrm{L}_1 (u_1 , 0)
\sim
\begin{pmatrix}
 -z_1 , & 1
\end{pmatrix}
\begin{pmatrix}
 1 & 1 \\
 z_1 & z_1
\end{pmatrix} =
\begin{pmatrix}
0 , & 0
\end{pmatrix}
$$
we conclude that it is equal to zero due to \eqref{qbasic}.

Then consider the  first diagonal matrix element in (\ref{qRL})
$
\begin{pmatrix}
 1 ,& 0
\end{pmatrix}
\mathrm{R}^2(u_1 ,u_2 | 0 ) \, \mathrm{L}_1 (u_1 , u_2)
\begin{pmatrix}
1 \\ z_2
\end{pmatrix}
$.
We are going to show that it is proportional to the
operator $\mathrm{R}^2$ with shifted arguments.
From \eqref{qbasic}
one gets
$$
\mathrm{R}^2(u) \, \mathrm{L}_1 (u_1 , u_2)
\begin{pmatrix}
1 \\ z_2
\end{pmatrix}
=
\mathrm{L}_1 ( u_1 , 0 )
\begin{pmatrix}
1 & 1 \\ z_2 q^{-u_2} & z_2 q^{u_2}
\end{pmatrix}
\begin{pmatrix}
q^{z_2 \partial_2} \, \mathrm{R}^2(u) \, q^{-z_2 \partial_2} \\ 0
\end{pmatrix},
$$
where we use the shorthand notation $\mathrm{R}^2(u) = \mathrm{R}^2(u_1,u_2|0)$.
Further we note that due to the previous formula, the second
factorization formula (\ref{R2}) for $\mathrm{R}^2(u)$
\begin{equation} \label{R2twist}
q^{z_2 \partial_2} \,\mathrm{R}^2(u) \, q^{-z_2 \partial_2} =
\mathrm{S}^3 (u_1) \, q^{z_2 \partial_2} \,
\mathrm{S}^2( u_2 ) \, q^{-z_2 \partial_2} \, \mathrm{S}^3 ( u_2 - u_1 )\,,
\end{equation}
and the intertwining relation (\ref{S3})
%$$
%\mathrm{S}_3(u_1) \, \mathrm{L}_1(0,u_1) =
%\mathrm{L}_1(u_1,0) \, \mathrm{S}_3(u_1)
%$$
the wanted matrix element takes the form
$$
\mathrm{S}^3(u_1) \cdot
\underline{
\begin{pmatrix}
 1 , & 0
\end{pmatrix}
\mathrm{L}_1(0,u_1)
\begin{pmatrix}
1 \\ z_2 q^{-u_2}
\end{pmatrix} } \cdot
q^{z_2 \partial_2} \, \mathrm{S}^2( u_2 ) \, q^{-z_2 \partial_2} \,
\mathrm{S}^3 ( u_2 - u_1 )\,.
$$
Taking into account (\ref{Lq})
we notice that the underlined matrix composition in the latter formula
takes the form
$$
\begin{pmatrix}
 1 , & 1
\end{pmatrix}
\begin{pmatrix}
q^{z_1 \partial_{z_1}+1} & 0\\
0 & q^{-z_1 \partial_{z_1}-1}
\end{pmatrix}
\begin{pmatrix}
 1 \\
 -1
\end{pmatrix} \cdot
\begin{pmatrix}
1-\frac{z_2}{z_1}q^{-u_2}
\end{pmatrix}
$$
and due to the recurrence formula (\ref{useful3}) for the argument shift in
$\mathrm{S}^2$ we obtain that the matrix element is equal to
$$
\mathrm{S}^3(u_1) \, \frac{1}{z_1} \left( q^{z_1 \partial_{z_1}} - q^{-z_1 \partial_{z_1}} \right) \,
\mathrm{S}^2( u_2 +1 ) \, \mathrm{S}^3 ( u_2 - u_1 )\,.
$$
In the final step we  use the  recurrence formula for the operator
$\mathrm{S}^3$ (\ref{rec}) to shift its argument
and obtain the first diagonal matrix element
$$
- q^{- \frac{1}{2} }\,
  \mathrm{S}^3(u_1 + 1)\,
  \mathrm{S}^2(u_2 + 1)\,
  \mathrm{S}^3(u_2 - u_1)\,.
$$

Consider the second diagonal matrix element in \eqref{qRL}:
$
\begin{pmatrix}
 -z_1 , & 1
\end{pmatrix}
\mathrm{R}^2(u_1 , u_2 | 0 ) \, \mathrm{L}_1 (u_1 , u_2)
\begin{pmatrix}
0 \\ 1
\end{pmatrix}
$.
Due to \eqref{qbasic2} it  is equal to
\begin{equation} \label{eqB}
\frac{1}{z_2 (q^{v_2} - q^{-v_2})}
\begin{pmatrix}
  -z_1 , & 1
\end{pmatrix}
\mathrm{L}_1 (u_1 , v_2)
\begin{pmatrix}
1 & 1 \\
z_2 q^{-u_2} & z_2 q^{u_2}
\end{pmatrix}
\begin{pmatrix}
- q^{z_2 \partial_2} \, \mathrm{R}^2(u_1 , u_2 | v_2) \, q^{-z_2 \partial_2}
\\
q^{-z_2 \partial_2} \, \mathrm{R}^2(u_1 , u_2 | v_2) \, q^{z_2 \partial_2}
\end{pmatrix},
\end{equation}
where we need to take  $v_2 \to0$ carefully.
%Further we will compute (\ref{eq:B})
%using either recurrence relation (\ref{eq:recurrence1}) or (\ref{eq:recurrence2}).
%\begin{itemize}
%\item
In this limit we have
$$
\frac{1}{z_2 (q^{v_2} - q^{-v_2})}
\begin{pmatrix}
  -z_1 , & 1
\end{pmatrix}
\begin{pmatrix}
1 & 1 \\ z_1 q^{-v_2} & z_1 q^{v_2}
\end{pmatrix} \to
\frac{1}{2} \frac{z_1}{z_2}
\begin{pmatrix}
-1 , & 1
\end{pmatrix}
$$
and consequently taking into account the factorization
of the $\mathrm{L}$-operator (\ref{Lq}),
performing the argument shift in $\mathrm{S}^2$
by means of \eqref{useful6}, \eqref{useful7}
$$
\begin{pmatrix}
1 & 1 \\
z_2 q^{-u_2} & z_2 q^{u_2}
\end{pmatrix}
\begin{pmatrix}
- q^{z_2\partial_{z_2}} \mathrm{S}^2(u_2) q^{-z_2\partial_{z_2}} \\
q^{-z_2\partial_{z_2}} \mathrm{S}^2(u_2) q^{z_2\partial_{z_2}}
\end{pmatrix} =
z_2 (q^{u_2}-q^{-u_2})
\begin{pmatrix}
1 \\ z_1
\end{pmatrix}
\mathrm{S}^2(u_2-1)
$$
we see that in the limit $v_2 \to 0$ (\ref{eqB}) takes the form
$$
\frac{1}{2} \frac{z_1}{z_2}
\begin{pmatrix}
-1 , & 1
\end{pmatrix}
\underline{
\begin{pmatrix}
q^{z_1 \partial_{z_1} + 1} & 0 \\
0 & q^{-z_1 \partial_{z_1} - 1}
\end{pmatrix}
\begin{pmatrix}
q^{u_1} & -z_1^{-1} \\
-q^{-u_1} & z^{-1}_1
\end{pmatrix}
\mathrm{S}^3(u_1)
\begin{pmatrix}
1 \\ z_1
\end{pmatrix} }
z_2 (q^{u_2} - q^{-u_2}) \, \mathrm{S}^2(u_2-1)\,
\mathrm{S}^3(u_2-u_1).
$$
In the previous formula the underlined
expression can be transformed using the matrix recurrent relation (\ref{rec2})
and finally we find that the wanted matrix element is equal to
$$
- q^{ \frac{1}{2} } ( q^{u_1} - q^{-u_1} )( q^{u_2} - q^{-u_2} ) \,
\mathrm{S}^3(u_1 - 1) \, \mathrm{S}^2(u_2 - 1) \, \mathrm{S}^3(u_2 - u_1)\,.
$$
Now the calculation of the second diagonal matrix element
can be  completed in a similar way using the other recurrent formula (\ref{rec1})
instead of (\ref{rec2}).

The result of the computation is the following matrix \eqref{qRL} (cf. (\ref{cornerstone}))
\begin{equation*}
%\begin{array}{c}
\begin{pmatrix}
1 & 0 \\
-z_1 & 1
\end{pmatrix}
\mathrm{R}^2(u) \mathrm{L}_1 (u_1 ,u_2)
\begin{pmatrix}
1 & 0 \\
z_2 & 1
\end{pmatrix}
=
\begin{pmatrix}
- q^{- \frac{1}{2} } \, \mathrm{R}^2(u + 1) & \cdots \\
0 & - q^{ \frac{1}{2} } ( q^{u_1} - q^{-u_1} )( q^{u_2} - q^{-u_2} ) \, \mathrm{R}^2(u - 1)
\end{pmatrix}.
%\end{array}
\end{equation*}
 As it has been explained in Section 2 we readily obtain the
Baxter equation (cf. (\ref{BE}))
\begin{equation} \label{qBE}
\mathrm{t}(u)\, \mathrm{Q}_i(u) = (-)^{N} q^{-\frac{N}{2}}\, \mathrm{Q}_i(u+1) +
(-)^{N} q^{\frac{N}{2}}\, \Delta^N(u_1,u_2)\,\mathrm{Q}_i(u-1)\;\;,\;\; i = 1,2
\end{equation}
where $\Delta(u_1,u_2) = ( q^{u_1} - q^{-u_1} )( q^{u_2} - q^{-u_2} )$ is symmetric.
The Baxter equation for $\mathrm{Q}_1$ follows from the fact that
both $\mathrm{Q}$-operators are connected by similarity  transformation (\ref{QTau}).
%%%%%%%%%%%%%%%%%%%%%%%%%%%%%%%%%%%%%%%%%%%%%%%%%%%%%%%%%%%%%%%%%%%%%%%%%%%%%%%%%%%%%%%%%%%%%%%%%%%%%%%%%%%%%%%%%%%
%%%%%%%%%%%%%%%%%%%%%%%%%%%%%%%%%%%%%%%%%%%%%%%%%%%%%%%%%%%%%%%%%%%%%%%%%%%%%%%%%%%%%%%%%%%%%%%%%%%%%%%%%%%%%%%%%%%%
%%%%%%%%%%%%%%%%%%%%%%%%%%%%%%%%%%%%%%%%%%%%%%%%%%%%%%%%%%%%%%%%%%%%%%%%%%%%%%%%%%%%%%%%%%%%%%%%%%%%%%%%%%%%%%%%%%%
%%%%%%%%%%%%%%%%%%%%%%%%%%%%%%%%%%%%%%%%%%%%%%%%%%%%%%%%%%%%%%%%%%%%%%%%%%%%%%%%%%%%%%%%%%%%%%%%%%%%%%%%%%%%%%%%%%%
%%%%%%%%%%%%%%%%%%%%%%%%%%%%%%%%%%%%%%%%%%%%%%%%%%%%%%%%%%%%%%%%%%%%%%%%%%%%%%%%%%%%%%%%%%%%%%%%%%%%%%%%%%%%%%%%%%%%
%%%%%%%%%%%%%%%%%%%%%%%%%%%%%%%%%%%%%%%%%%%%%%%%%%%%%%%%%%%%%%%%%%%%%%%%%%%%%%%%%%%%%%%%%%%%%%%%%%%%%%%%%%%%%%%%%%%
\subsection{Explicit action of the trigonometric $\mathrm{Q}$-operator}
\label{qExplQ}
%%%%%%%%%%%%%%%%%%%%%%%%%%%%%%%%%%%%%%%%%%%%%%%%%%%%%%%%%%%%%%%%%%%%%%%%%%%%%%%%%%%%%%%%%%%%%%%%%%%%%%%%%%%%%%%%%%%
Here we are going to establish an explicit formula for the operator
$\mathrm{Q}_2$ (\ref{Q2})
%%which is constructed out of several copies of local operator $\mathrm{R}^2$.
 by computing its action  on the generating function
of the symmetry algebra representation in the quantum space of the spin
chain states.
It turns out that due to (\ref{trA}) it is sufficient to
know the action  by $\mathrm{R}^2$
on the generating function of the representation in one site of the spin chain.
The latter observation simplifies the task considerably reducing
the global problem to a local one.
Moreover the local problem can be solved up to a constant without any
reference to an explicit expression for the operator $\mathrm{R}^2$ (like (\ref{R2}))
but using only the its  defining relation (\ref{R2LL}).
We need the explicit expression for $\mathrm{R}^2$ only in the last step
for fixing a constant multiplier.
%%:
%\begin{equation*}
%\mathrm{R}^{2}(u_1,u_2|v_2) \, \mathrm{L}_1(u_1,u_2) \, \mathrm{L}_2(v_1,v_2) = \
%\mathrm{L}_1(u_1,v_2) \, \mathrm{L}_2(v_1,u_2) \, \mathrm{R}^{2}(u_1,u_2|v_2)
%\end{equation*}
Let us remind that in (\ref{R2LL}) the involved $\mathrm{L}$-operators act
 in two different quantum spaces. Taking into account (\ref{Yangian})
we see that $\mathrm{L}_1 \cdot \mathrm{L}_2$ defines a co-product like structure
on the quantum algebra $\mathcal{U}_q(s\ell_2)$.
Further in (\ref{R2LL}) we  shift the  four parameters
$u_i \to u_i + \lambda$, $v_i \to v_i + \lambda$ ($i=1,2$)
which does not change the operator $\mathrm{R}^2$,
extract the  matrix element below diagonal,
consider the asymptotics $\lambda \to +\infty$ and obtain the following relation
\begin{equation} \label{R2DeltaS+}
\begin{array}{c}
\mathrm{R}^{2}(u_1,u_2|v_2)
\left[ \mathbf{S}^{+}_1 (u_1-{u_2}+1) \, \mathbf{K}_2(-v_1-1)
  + \mathbf{K}^{-1}_1({u_2}) \, \mathbf{S}^{+}_2 (v_1-{{v_2}}+1)\right] = \\ [0.3 cm]
\left[ \mathbf{S}^{+}_1 (u_1-{{v_2}}+1) \, \mathbf{K}_2(-v_1-1)
  + \mathbf{K}^{-1}_1({{v_2}}) \, \mathbf{S}^{+}_2 (v_1-{u_2}+1)\right]
\mathrm{R}^{2}(u_1,u_2|v_2)
\end{array}
\end{equation}
where we use the shorthand notations
$
\mathbf{S}^{+}(a) = z [ z \partial_z + a] \;\; ;\; \; \mathbf{K}(a) = q^{z \partial_z - a}
$.
In (\ref{R2DeltaS+})  the following co-product like expression appears
depending on some spectral parameters
$$
(\Delta \mathbf{S}^{+})(a,b,c,d) =
\mathbf{S}^{+}_1 (a) \, \mathbf{K}_2(-b) + \mathbf{K}^{-1}_1(c) \, \mathbf{S}^{+}_2(d)\,.
$$
Then from (\ref{R2DeltaS+}) it follows that
\begin{equation} \label{R2FDeltaS+}
\begin{array}{c}
\mathrm{R}^{2}(u_1,u_2|v_2) \,
\mathcal{F}\left( \lambda (\Delta \mathbf{S}^{+})
(u_1-{u_2}+1 , v_1+1 , {u_2} , v_1-{{v_2}}+1) \right) \cdot 1 = \\ [0.3 cm]
= \mathcal{F}\left( \lambda (\Delta \mathbf{S}^{+})
( u_1-{{v_2}}+1 , v_1+1 , {{v_2}} , v_1-{u_2}+1) \right)
\mathrm{R}^{2}(u_1,u_2|v_2) \cdot 1
\end{array}
\end{equation}
where $\mathcal{F}$ is any function. Further we choose it to be the $q$-series
\begin{equation} \label{F}
\mathcal{F}(x) = \sum_{k\geq 0} \frac{ q^{\frac{n(n+1)}{2}} }{(q^2;q^2)_n} (q^{-1}-q)^n x^n\,.
\end{equation}
In order to calculate $\mathcal{F}\left(\lambda (\Delta \mathcal{S}^{+})\right) \cdot 1$
we need the following formula
$$
\left((\Delta \mathbf{S}^{+})(a,b,c,d)\right)^n \cdot 1 = \sum_{k=0}^n
\frac{(q^{2a};q^2)_k (q^{2d};q^2)_{n-k} (q^2;q^2)_n}{(q^{-1}-q)^n (q^2;q^2)_k (q^2;q^2)_{n-k}}
q^{(b-a) k + (c-d)(n-k) -\frac{n(n-1)}{2}} z^k_1 z^{n-k}_2
$$
which can be established easily by induction.
Then by means of the $q$-binomial formula (\ref{qBinom})
one obtains straightforwardly\footnote{Here and below we adopt the notation
$(a,b,c,\cdots;q^2)=(a;q^2)(b;q^2)(c;q^2)\cdots$.}
$$
\mathcal{F}\left(\lambda (\Delta \mathbf{S}^+)(a,b,c,d)\right)\cdot 1 =
\frac{(q^{1+a+b}\lambda z_1 , q^{1+c+d}\lambda z_2;q^2)}
     {(q^{1-a+b}\lambda z_1 , q^{1+c-d}\lambda z_2;q^2)}\,.
%%\frac{(q^{1+a+b}\lambda z_1;q^2)}{(q^{1-a+b}\lambda z_1;q^2)}
%%\frac{(q^{1+c+d}\lambda z_2;q^2)}{(q^{1+c-d}\lambda z_2;q^2)}
$$
Thus we see that $\mathcal{F}\left(\lambda (\Delta \mathbf{S}^{+})\right) \cdot 1$
is a product of two factors, the first being a function of $z_1$
and the second being a function of $z_2$.
This  happens due to the special choice of the function $\mathcal{F}$
(\ref{F}). This factorization
 is crucial for our purpose, because by the commutativity
of $\mathrm{R}^2$ with $z_2$ the result  (\ref{R2FDeltaS+})
can be rewritten as
\begin{equation} \label{qR2GenFun}
\mathrm{R}^2_{12}(u)
\frac{(q^{-2\ell}\lambda z_1;q^2)}{(q^{+2\ell}\lambda z_1;q^2)} = c \cdot
\frac{(q^{+u-\ell}\lambda z_1  , q^{-u-\ell}\lambda z_2 ;q^2)}
     {(q^{-u+\ell}\lambda z_1 , q^{+u+\ell}\lambda z_2 ;q^2)}\,,
%%\frac{(q^{u-\ell}\lambda z_1;q^2)}{(q^{-u+\ell}\lambda z_1;q^2)}
%%\frac{(q^{-u-\ell}\lambda z_2;q^2)}{(q^{u+\ell}\lambda z_2;q^2)}
\end{equation}
where we choose $v_1 = -2, v_2 = 0$ and calculate the
constant $c=\mathrm{R}^2(u) \cdot 1$
using one of the possible forms of $\mathrm{R}^2$ (\ref{4Fact'})
$$
c = \mathrm{R}^2(u) \cdot 1 =
q^{-u_1 u_2 + \frac{u_2^2}{2}}
\frac{(q^{2+2 u_1-2 u_2};q^2)}{(q^{2+2 u_1};q^2)}\,.
$$
With this form of the action of $\mathrm{R}^2$ we proceed to the action of
$\mathrm{Q}_2$. At this point it is convenient to
 renormalize the operator $\mathrm{Q}_2$:
$\mathrm{Q}_2(u) \to c^{-N}\cdot\mathrm{Q}(u)$, such that the Baxter
equation takes the form
$$
\mathrm{t}(u)\, \mathrm{Q}(u) = \Delta_{+}(u)\, \mathrm{Q}(u+1) +
\Delta_{-}(u)\,\mathrm{Q}(u-1)\,,
$$
where $\Delta_{\pm}(u) = (q^{u\mp\ell}-q^{-u\pm\ell})^N$. Now we see that
 due to (\ref{trA})
the Baxter operator $\mathrm{Q}(u)$ acts on the generating function
of spin chain states as follows
\begin{equation} \label{qQ}
\mathrm{Q}(u) \cdot \prod_{i=1}^{N}
\frac{(q^{-2\ell} \, \lambda_i \, z_i;q^2)}{(q^{+2\ell} \, \lambda_i \, z_i;q^2)} =
\prod_{i=1}^{N}
\frac{( q^{-u-\ell} \, \lambda_i \, z_i , q^{+u-\ell} \, \lambda_{i+1} \, z_i ;q^2)}
     {( q^{+u+\ell} \, \lambda_i \, z_i , q^{-u+\ell} \, \lambda_{i+1} \, z_i
;q^2)}\,,
\end{equation}
where as usual we assume cyclicity, $N+1 \equiv 1$.
%%Let us stress that the above calculation is based on the
%%intertwining relation (\ref{R2LL}) which defines a 'co-product'
%%structure, commutativity of $\mathrm{R}^2$ with $z_2$,
%%trivial action of $\mathrm{R}^2$ on a constant
%%and the possibility to choose.

In Appendix \ref{altern}
we propose an alternative proof of (\ref{qR2GenFun})
which uses only the Coxeter relation (\ref{S3S2S3})
and is very similar to the derivation of the
analogous formula in the elliptic case in Subsection \ref{ellipExplQ}.

Thus for a spin chain with $q$-deformed symmetry algebra
we have constructed a pair of Baxter $\mathrm{Q}$-operators,
proved the corresponding Baxter equation and found an explicit formula for one of them.
Such operators respects commutativity (\ref{commut}) and factorization (\ref{Factor})
property
as it has been explained in Section 2 on the basis of  general arguments.
%%%%%%%%%%%%%%%%%%%%%%%%%%%%%%%%%%%%%%%%%%%%%%%%%%%%%%%%%%%%%%%%%%%%%%%%%%%%%%%%%%%%%%%%%%%%%%%%%%%%%%%%%%%%%%%%%%%
%%%%%%%%%%%%%%%%%%%%%%%%%%%%%%%%%%%%%%%%%%%%%%%%%%%%%%%%%%%%%%%%%%%%%%%%%%%%%%%%%%%%%%%%%%%%%%%%%%%%%%%%%%%%%%%%%%%%
%%%%%%%%%%%%%%%%%%%%%%%%%%%%%%%%%%%%%%%%%%%%%%%%%%%%%%%%%%%%%%%%%%%%%%%%%%%%%%%%%%%%%%%%%%%%%%%%%%%%%%%%%%%%%%%%%%%
\section{Elliptic deformation case}
%%Sklyanin symmetry algebra
\setcounter{equation}{0}
\label{ellip}
%%%%%%%%%%%%%%%%%%%%%%%%%%%%%%%%%%%%%%%%%%%%%%%%%%%%%%%%%%%%%%%%%%%%%%%%%%%%%%%%%%%%%%%%%%%%%%%%%%%%%%%%%%%%%%%%%%%
%%%%Elliptic deformation. Notations and ingredients of the $\mathrm{R}$-operator construction.
\subsection{Elliptic $\mathrm{L}$-operator and parameter permutation operators}
%%%%%%%%%%%%%%%%%%%%%%%%%%%%%%%%%%%%%%%%%%%%%%%%%%%%%%%%%%%%%%%%%%%%%%%%%%%%%%%%%%%%%%%%%%%%%%%%%%%%%%%%%%%%%%%%%%%
Now we proceed to the most intricate example of deformation
 applying the general arguments of Section 2. The
elliptically deformed $s\ell_2$  symmetry
algebra has been introduced in \cite{Skl82,Skl83} and is called Sklyanin algebra.
Its defining relations are equivalent to (\ref{Yangian}) where the
numerical $\mathcal{R}$-matrix is due to Baxter and
appeared first in his solution of the eight-vertex model \cite{Baxter}.

Further we are interested in infinite-dimensional representations of the algebra
in the space of meromorphic even functions of one complex variable \cite{Skl83}. It is realized by
second order finite difference operators which are
constructed out of Jacobi theta functions (see Appendix \ref{SpFun}),
depend on two deformation
parameters $\eta$, $\tau$ and on spin $\ell$ being an arbitrary complex number.
%%%%Substituting this operator representation of the
%%%%algebra generators in (\ref{L})
%%%%one can obtain the factorized representation for
%%%%$\mathrm{L}$-operator \cite{ParamPerm}
Choosing this operator representation of the
algebra generators
one can obtain the factorized form for the 
$\mathrm{L}$-operator \cite{Zab,Zab11,ParamPerm}
\begin{equation} \label{Lellip}
\mathrm{L}(u_1,u_2) =
\frac{1}{\theta_1(2 z)}\,
\mathrm{M}(z\mp u_2)\,
\begin{pmatrix}
e^{\eta \partial_z} & 0 \\
0 & e^{-\eta \partial_z}
\end{pmatrix}\,
\mathrm{N}(z\mp u_1)
\end{equation}
where we denote matrices involving theta functions as follows
\begin{equation} \label{M;N}
\mathrm{M}(a\mp b)=
\begin{pmatrix}
(a-b)_3 & -(a+b)_3 \\
-(a-b)_4 & (a+b)_4
\end{pmatrix},\;\; \;\;
\mathrm{N}(a\mp b)=
\begin{pmatrix}
(a+b)_4 & (a+b)_3 \\
(a-b)_4 & (a-b)_3
\end{pmatrix},
\end{equation}
and specify the relation between two sets of spectral parameters
$(u_1,u_2)$ and $(u,\ell)$ according to (\ref{ul}) as
\begin{equation} \label{uEl}
u_1 = \frac{u}{2} - \eta  \ell - \eta \;\;, \;\;
u_2 = \frac{u}{2} + \eta  \ell \,.
\end{equation}
All notations and properties of theta functions relevant for us
are collected in Appendix \ref{SpFun}.
Let us note that by (\ref{thetatheta})
the matrices $\mathrm{M}$ and $\mathrm{N}$ (\ref{M;N}) are inverse to each other
\begin{equation} \label{MN}
\mathrm{M}(a \mp b) \, \mathrm{N}(a \mp b) =
2 \, \theta_1(2 a) \, \theta_1(2 b) \cdot \II\,.
\end{equation}

This $\mathrm{L}$-operator (\ref{Lellip}) has been used
in \cite{ParamPerm}
to construct the elementary permutation operators $\mathrm{S}^i(a)$ ($i=1,2,3$)
and corresponding general $\mathrm{R}$-operator respecting (\ref{RLL}).
However  it has been demonstrated in \cite{EllDerSp} that it is more
convenient  to work with a slightly modified version of (\ref{Lellip}).
Due to invariance of Baxter's $\mathcal{R}$-matrix:
$\sigma_3\otimes\sigma_3 \, \mathcal{R}(u) = \mathcal{R}(u)\, \sigma_3\otimes\sigma_3$ ,
we conclude that the $\mathrm{L}$-operator multiplied by
Pauli matrix $\sigma_3$ on the left
$\sigma_3\,\mathrm{L}(u)$ solves (\ref{Yangian}) as well
and, consequently, can be substituted as $\mathrm{L}$-operator.
This transformation, $\mathrm{L}(u)\to \sigma_3\, \mathrm{L}(u)$,
corresponds to a certain automorphism of the Sklyanin algebra.
%%It  leads also to the functions in the representation space
%%$\mathbb{V}_{\ell}$ to be of the form $e^{-\frac{\pi i}{\eta} z^2}f(z)$
%%where $f(z)$ is an arbitrary meromorphic function \cite{KriZab}.
Further we solve the $\mathrm{RLL}$-relation (\ref{RLL})
in the form
\begin{equation*} %%\label{RLsigmaL}
\mathrm{R}_{12}(u_1,u_2|v_1,v_2) \, \sigma_3 \, \mathrm{L}_1 (u_1 , u_2) \,\sigma_3\, \mathrm{L}_2 (v_1 , v_2) =
\sigma_3\,\mathrm{L}_1 (v_1 , v_2) \,\sigma_3\, \mathrm{L}_2 (u_1 , u_2) \; \mathrm{R}_{12}(u_1,u_2|v_1,v_2)
\end{equation*}
where the $\mathrm{L}$-operator is given in (\ref{Lellip}).
The same substitution $\mathrm{L}(u) \to \sigma_3 \, \mathrm{L}(u)$ should be done also
in the other formulae in Section 2:
(\ref{S1}), (\ref{S2LL}), (\ref{S3}), (\ref{R1LL}), (\ref{R2LL}).

Now we are ready to present the operators of elementary permutations
$\mathrm{S}^1$, $\mathrm{S}^2$ and $\mathrm{S}^3$.
Their construction and consequently the
integrability structure of the spin chain is based on the
elliptic gamma function $\Gamma(z|\tau,2\eta)$.
Further we use for it the shorthand notation $\Gamma(z)$.
Its definition and properties relevant for us
a collected in Appendix \ref{SpFun}.
%%%%%%%%%%%%%%%%%%%%%%%%%%%%%%%%%%%%%%%%%%%%%%%%%%%%%%%%%%%%%%%%%%%%%%%%%%%%%%%%%%%%%%%
%%The function which plays the main role
%%in the further study of elliptic deformation is elliptic
%%gamma-function
%%Its peculiar properties underlie the integrable structure.
%%We denote it in the following as $\Gamma(z) = \Gamma(z|\tau,2\eta)$.
%%To establish integrable structure
%%we need to find the general $\mathrm{R}$-matrix
%%acting in the tensor product of two infinite-dimensional
%%representations of the Sklyanin algebra.
%%It can be achieved solving
%%$\mathrm{RLL}$-relation with
%%elliptic $\mathrm{L}$-operator
%%%%%%%%%%%%%%%%%%%%%%%%%%%%%%%%%%%%%%%%%%%%%%%%%%%%%%%%%%%%%%%%%%%%%%%%%%%%%%%%%%%%%%%%%%%%
%%Let us note that operator (\ref{LFact}) multiplied by Pauli matrix $\sigma_3$
%%on the left is also $\mathrm{L}$-operator since it respects $\mathrm{RLL}$-relation
%%with numerical $\mathrm{R}$-matrix of the $XYZ$ spin chain.
%%%%%%%%%%%%%%%%%%%%%%%%%%%%%%%%%%%%%%%%%%%%%%%%%%%%%%%%%%%%%%%%%%%%%%%%%%%%%%%%%%%%%%%%%%%%%%%%%%%
%%Further we cite two basic ingredients in the construction of the general $\mathrm{R}$-operator
%%%%%%%%%%%%%%%%%%%%%%%%%%%%%%%%%%%%%%%%%%%%%%%%%%%%%%%%%%%%%%%%%%%%%%%%%%%%%%%%%%%%%%%%%%%%%%%%%%%
\begin{itemize}
\item
The operator $\mathrm{S}^2$ acts nontrivially in the both quantum spaces
and is defined by the operator relation (\ref{S2LL})
where $\mathrm{L}(u) \to \sigma_3 \, \mathrm{L}(u)$ (\ref{Lellip}).
%%\begin{equation} \label{S2LsigmaL}
%%\mathrm{S}_2(u_1-v_2) \, \sigma_3 \, \mathrm{L}_1 (u_1 , u_2) \,\sigma_3\, \mathrm{L}_2 (v_1 , v_2) =
%%\sigma_3\,\mathrm{L}_1 (v_2 , u_2) \,\sigma_3\, \mathrm{L}_2 (v_1 , u_1) \; \mathrm{S}_2(u_1-v_2)
%%\end{equation}
One of the possible solutions of the this relation
which is suitable for our purposes
has the form\footnote{Here and below we adopt the notation
$\Gamma(\mp a \mp b)=\Gamma(a+b)\Gamma(-a+b)\Gamma(a-b)\Gamma(-a-b)$.} \cite{EllDerSp}
\begin{equation} \label{ellipS2}
\begin{array}{c}
\mathrm{S}^2(a) = \Gamma(\mp z_1 \mp z_2 +a + \eta + \frac{\tau}{2})\,.
\end{array}
\end{equation}
A similar expression appeared in \cite{ParamPerm} with some additional
exponentials and without the shift by $\frac{\tau}{2}$.  
Exactly the same expression was used in \cite{BS} for
the formulation of the star-triangle relation.
\item
The infinite-dimensional representations of the Sklyanin algebra
parameterized by $\ell$ and $-\ell-1$ are equivalent since the Casimir
operators take coinciding numerical values for both representations.
The corresponding intertwining operator $\mathrm{W}(\eta(2\ell+1))$ (\ref{WS})
can be realized as an integral operator \cite{EllDerSp}
on the space of even functions
\begin{equation} \label{ellipW}
\mathrm{W}(a) \, \Phi(z) =
\int^{1}_{0} \mathrm{d} x \, \mu(x) \frac{e^{-\frac{\pi i}{\eta}(z^2+x^2)}}{\Gamma(-2 a)}\,
\Gamma(\mp z \mp x -a) \, \Phi(x) \;\;;\;\; \mu(x) =
\frac{\mathrm{C}\,e^{\frac{2 \pi i}{\eta} x^2}}{\Gamma(\mp 2 x)}\,,
\end{equation}
where $\mu(x)$ is the integration measure
and the  constant is
$\mathrm{C} = \frac{1}{2} \,(e^{4 \pi i \eta};e^{4 \pi i \eta}) \,(e^{2 \pi i \tau};e^{2 \pi i \tau})$.
It is remarkable that a similar operator 
was used in \cite{spi:bailey} for the construction 
of the integral Bailey transformation. 
The formula which is equivalent to 
the binary relation of permutation group
$\mathrm{W}(a)\mathrm{W}(-a)=\II$ was proved in 
\cite{spi-war:inversions} in the context
of integral Bailey transformation. 
Unlike the $q$-deformed case
the exponential property is missing.
The operators $\mathrm{S}^1(a)$ and $\mathrm{S}^3(a)$ (\ref{S1}), (\ref{S3})
are two copies of the intertwining operator $\mathrm{W}(a)$ which act
nontrivially in the
second and the first quantum spaces, respectively (\ref{S1S2S3}).
\end{itemize}

Finally, the  indicated operators $\mathrm{S}^1, \mathrm{S}^2, \mathrm{S}^3$
respect Coxeter relations
(\ref{S1S2S1}), (\ref{S3S2S3}) \cite{EllDerSp}
that is the direct consequence of
the elliptic beta integral evaluation formula by V.~Spiridonov \cite{Sp}.

For comparison to \cite{ParamPerm} we remark that the elementary permutation
operators has been constructed there using the $\mathrm{L}$-operator (\ref{Lellip}).
The resulting operator $\mathrm{S}^2$ is not symmetric under
$z_1 \leftrightarrow z_2$ contrary to (\ref{ellipS2}). Further, the
intertwining operator $\mathrm{W}$ of Sklyanin algebra representations
has been taken in the form of the operator series
constructed by A.~Zabrodin. Despite of the fact that the existence of this series
in the infinite-dimensional representation case is elusive
the  Coxeter relations has been proven
by means of formal manipulations on the  operator series
using the  Frenkel-Turaev summation formula \cite {FT}.
%%%%%%%%%%%%%%%%%%%%%%%%%%%%%%%%%%%%%%%%%%%%%%%%%%%%%%%%%%%%%%%%%%%%%%%%%%%%%%%%%%%%%%%%%%%%%%%%%%%%%%%%%%%%%%
%%Operator $\mathrm{W}(2\ell+1)$ intertwines two representations of the Sklyanin algebra
%%with parameters $\ell$ and $-\ell-1$ that can be stated as
%%$$
%%\mathrm{W}(u_2-u_1)\,\mathrm{L}(u_1,u_2) = \mathrm{L}(u_2,u_1)\,\mathrm{W}(u_2-u_1)
%%$$
%%This operator can be represented as an integral operator
%%Further we denote $\mathrm{S}_3(a) = \mathrm{W}_1(a)$ an intertwiner in the first
%%quantum space
%%$$
%%\mathrm{S}_3(u_2-u_1) \, \mathrm{L}(u_1,u_2) = \mathrm{L}(u_2,u_1)\,\mathrm{S}_3(u_2-u_1)
%%$$
%%and $\mathrm{S}_1(a) = \mathrm{W}_2(a)$
%%an intertwiner in the second quantum space
%%$$
%%\mathrm{S}_1(v_2-v_1) \, \mathrm{L}(v_1,v_2) = \mathrm{L}(v_2,v_1)\,\mathrm{S}_1(v_2-v_1)
%%$$
%%%%%%%%%%%%%%%%%%%%%%%%%%%%%%%%%%%%%%%%%%%%%%%%%%%%%%%%%%%%%%%%%%%%%%%%%%%%%%%%%%%%%%%%%%%%%%%%%%%%%%%%%%%%%%%%%%%
%%%%%%%%%%%%%%%%%%%%%%%%%%%%%%%%%%%%%%%%%%%%%%%%%%%%%%%%%%%%%%%%%%%%%%%%%%%%%%%%%%%%%%%%%%%%%%%%%%%%%%%%%%%%%%%%%%%%
%%%%%%%%%%%%%%%%%%%%%%%%%%%%%%%%%%%%%%%%%%%%%%%%%%%%%%%%%%%%%%%%%%%%%%%%%%%%%%%%%%%%%%%%%%%%%%%%%%%%%%%%%%%%%%%%%%%
\subsection{Elliptic recurrence relations}
\label{recurrEl}
%%%%%%%%%%%%%%%%%%%%%%%%%%%%%%%%%%%%%%%%%%%%%%%%%%%%%%%%%%%%%%%%%%%%%%%%%%%%%%%%%
Now we proceed to recurrence relations which
connect $\mathrm{S}^{i}(a)$ and $\mathrm{S}^{i}(a\pm \eta)$ ($i=1,2,3$).
As we shall see  their form
is very similar to the undeformed and the trigonometric case (\ref{qRecurr}).
%%Here we cite recurrence relations for intertwining operator $\mathrm{W}(a)$
%%of representations of the Sklyanin algebra

We start with  recurrence relations for the intertwining operator
$\mathrm{W}(a)$ (\ref{ellipW})
\begin{eqnarray}
\label{recaEl}
\mathrm{R}(\tau) \,e^{-\pi i \eta} (z)_i \mathrm{W}(a+\eta)
 =  \mathrm{W}(a) \, \frac{1}{\theta_1(2z)}
\left[ (z-a)_i \,e^{\eta \partial_z} -
(z+a)_i \, e^{-\eta \partial_z}
\right], \\ [0.2 cm]
%%\end{equation}
%%\begin{equation}
\label{recbEl}
\mathrm{R}(\tau) \,e^{-\pi i \eta} \, \mathrm{W}(a+\eta) \,(z)_i
 =  \frac{1}{\theta_1(2z)}
\left[ (z+a+\eta)_i \,e^{\eta \partial_z} -
(z-a-\eta)_i \, e^{-\eta \partial_z}
\right] \mathrm{W}(a)\,,
\end{eqnarray}
where $i=3,4$ and $\mathrm{R}(\tau)$ is a constant
(see Appendix \ref{SpFun}).
The second formula is a consequence of the first
one due to $\mathrm{W}(a) \mathrm{W}(-a) = \II$.
%% Since $\mathrm{W}(0)=\II$ at integer $l$ we obtain
%% a factorized representation for $\mathrm{W}(\eta l)$
%% $$
%% \begin{array}{c}
%% \mathrm{W}(\eta l) = \mathrm{R}^{-l}(\tau)\, e^{\pi i \eta l}\, \frac{1}{(z)^{l}_i} \cdot
%% \prod_{k=0}^{l-1} \frac{1}{\theta_1(2z)}
%% \left[(z-\eta k )_i \,e^{\eta \partial_z} -
%% (z+ \eta k)_i \, e^{-\eta \partial_z}\right]
%% = \\[0.5 cm] =
%% %\mathrm{W}(\eta l) =
%% \mathrm{R}^{-l}(\tau)\, e^{\pi i \eta l} \cdot
%% \prod_{k=0}^{l-1} \frac{1}{\theta_1(2z)}
%% \left[(z+\eta l-\eta k )_i \,e^{\eta \partial_z} -
%% (z- \eta l + \eta k)_i \, e^{-\eta \partial_z}\right]
%% \cdot \frac{1}{(z)^{l}_i}\,.
%% \end{array}
%% $$
%% %%It can be shown by induction that this
%% %%expression coincides with the one proposed
%% %%in \cite{} in the form of operator sum.
%% Expanding the latter formula we obtain a sum of finite-difference
%% operators which coincides with the expression for
%% intertwining operator at (half)-integer spin established in \cite{Zab98}.

 (\ref{recaEl}), (\ref{recbEl}) can be  rewritten
in  matrix form
\begin{eqnarray}
%%\begin{equation}
\label{recaaEl}
\mathrm{R}(\tau)\,
e^{-\pi i \eta}
\begin{pmatrix}
(z)_3 \\ -(z)_4\\
\end{pmatrix}
\mathrm{W}(a+\eta) =
\mathrm{W}(a)\,\frac{1}{\theta_1(2z)}\,
\mathrm{M}(z\mp a)
%%\begin{pmatrix}
%%(z-a)_3 & -(z+a)_3\\
%%-(z-a)_4 & (z+a)_4
%%\end{pmatrix}
\begin{pmatrix}
e^{\eta \partial_{z}}\\
e^{-\eta \partial_{z}}
\end{pmatrix},
%%\end{equation}
\\ [0.2 cm]
%%\begin{equation}
\label{recbbEl}
\mathrm{R}(\tau)\,
e^{-\pi i \eta}\,
\mathrm{W}(a+\eta)
\begin{pmatrix}
(z)_4 , & (z)_3\\
\end{pmatrix} =
\frac{1}{\theta_1(2z)}
\begin{pmatrix}
e^{\eta \partial_{z}} , &
-e^{-\eta \partial_{z}}
\end{pmatrix}
\mathrm{N}(z\mp a)
%%\begin{pmatrix}
%%(z+a)_4 & (z+a)_3\\
%%(z-a)_4 & (z-a)_3
%%\end{pmatrix}
\mathrm{W}(a)\,.
%%\end{equation}
\end{eqnarray}
%$$
%2 \, \mathrm{R}(\tau)^{-1}
%e^{\pi i \eta} \, \theta_1(2a)
%\begin{pmatrix}
%e^{\eta \partial_{z}}\\
%e^{-\eta \partial_{z}}
%\end{pmatrix}
%\mathrm{W}(a-\eta)
%= \begin{pmatrix}
%(z-a)_4 & (z-a)_3\\
%(z+a)_4 & (z+a)_3
%\end{pmatrix}
%\mathrm{W}(a)
%\begin{pmatrix}
%(z)_3 \\ -(z)_4\\
%\end{pmatrix}
%$$
Given the recurrence relations (\ref{recaaEl}) and (\ref{recbbEl})
it is easy to deduce
that $\mathrm{W}(a)$ is an intertwining operator, i.e. that
it respects:
$
\mathrm{W}(a) \, \mathrm{L}(0,a) =
\mathrm{L}(a,0) \, \mathrm{W}(a)
$.
Indeed due to (\ref{recaaEl}) and (\ref{ellipW}) the left hand side takes the form
$$
\mathrm{W}(a) \mathrm{L}(0,a) =
\mathrm{W}(a) \frac{1}{\theta_1(2z)}\,
\mathrm{M}(z\pm a)
\begin{pmatrix}
e^{\eta \partial_{z}} \\
e^{-\eta \partial_{z}}
\end{pmatrix} \otimes
\begin{pmatrix}
(z)_4 , & (z)_3
\end{pmatrix} =
c
\begin{pmatrix}
(z)_3 \\ -(z)_4
\end{pmatrix}
\mathrm{W}(a+\eta)
\otimes
\begin{pmatrix}
(z)_4 , & (z)_3
\end{pmatrix},
$$
where $c=\mathrm{R}(\tau)\,e^{-\pi i \eta}$,
and due to (\ref{recbbEl}) and (\ref{ellipW}) the right hand side
takes the same form
$$
\mathrm{L}(a,0) \mathrm{W}(a)=
\begin{pmatrix}
(z)_3 \\ -(z)_4
\end{pmatrix}
\otimes
\frac{1}{\theta_1(2z)}
\begin{pmatrix}
e^{\eta \partial_{z}} , &
- e^{-\eta \partial_{z}}
\end{pmatrix}
\mathrm{N}(z\pm a) \mathrm{W}(a) =
c
\begin{pmatrix}
(z)_3 \\ -(z)_4
\end{pmatrix}
\otimes
\mathrm{W}(a+\eta)
\begin{pmatrix}
(z)_4 , & (z)_3
\end{pmatrix}.
$$

The other two identities for $\mathrm{W}(a)$ (\ref{ellipW}) of interest in the
following are:
\begin{equation}
\label{recEl1}
- 2 \, \mathrm{R}(\tau)^{-1} \, e^{\pi i \eta} \, \theta_1(2a) \, \mathrm{W}(a-\eta)
\begin{pmatrix}
1 \\ 1
\end{pmatrix} =
\begin{pmatrix}
e^{\eta\partial_{z}} & 0 \\
0 & e^{-\eta \partial_{z}}
\end{pmatrix}
\mathrm{N}(z\mp a)
%%\begin{pmatrix}
%%(z+a)_4 & (z+a)_3 \\
%%(z-a)_4 & (z-a)_3
%%\end{pmatrix}
\mathrm{W}(a)
\begin{pmatrix}
(z)_3 \\ -(z)_4
\end{pmatrix},
\end{equation}
\begin{equation}
\label{recEl2}
- 2 \, \mathrm{R}(\tau)^{-1} \, e^{\pi i \eta} \,
\frac{\theta_1(2a)}{\theta_1(2z)} \, \mathrm{W}(a-\eta)
\begin{pmatrix}
1 \\ -1
\end{pmatrix}^T =
\begin{pmatrix}
(z)_4, & (z)_3
\end{pmatrix} \mathrm{W}(a)
\,\frac{1}{\theta_1(2z)}
\mathrm{M}(z\mp a)
%%\begin{pmatrix}
%%(z-a)_3 & -(z+a)_3 \\
%%-(z-a)_4 & (z+a)_4
%%\end{pmatrix}
\begin{pmatrix}
e^{\eta \partial_{z}} & 0 \\
0 & e^{-\eta \partial_{z}}
\end{pmatrix}.
\end{equation}
%%To see that the left hand side of the formula (\ref{recEl1}) is
%%proportional to the vector
%%$\begin{pmatrix} 1 \\ 1\end{pmatrix}$
%%we multiply the intertwining relation
%%$
%%\mathrm{W}(a) \, \mathrm{L}(0,a) =
%%\mathrm{L}(a,0) \, \mathrm{W}(a)
%%$
%%by the column
%%$\begin{pmatrix}
%% (z)_3 \\ -(z)_4
%%\end{pmatrix}$ on the right getting
%%$$
%%\mathrm{L}(a,0)\,\mathrm{W}(a)
%%\begin{pmatrix}
%% (z)_3 \\ -(z)_4
%%\end{pmatrix} =
%%\begin{pmatrix}
%%0 \\ 0
%%\end{pmatrix}
%%$$
%%which is equivalent to
%%$$
%%\begin{pmatrix}
%%1 ,& -1
%%\end{pmatrix}
%%\begin{pmatrix}
%%e^{\eta\partial_{z}} & 0 \\
%%0 & e^{-\eta \partial_{z}}
%%\end{pmatrix}
%%\begin{pmatrix}
%%(z+a)_4 & (z+a)_3 \\
%%(z-a)_4 & (z-a)_3
%%\end{pmatrix}
%%\mathrm{W}(a)
%%\begin{pmatrix}
%%(z)_3 \\ -(z)_4
%%\end{pmatrix}  = 0
%%$$
To see that the right hand side of  (\ref{recEl2})
is proportional to the row $\begin{pmatrix} 1, & -1\end{pmatrix}$
it is sufficient to multiply
$
\mathrm{W}(a) \, \mathrm{L}(0,a) =
\mathrm{L}(a,0) \, \mathrm{W}(a)
$ by the row
$\begin{pmatrix}
(z)_4 , & (z)_3
\end{pmatrix}$
from the left to obtain
$
\begin{pmatrix}
(z)_4 , & (z)_3
\end{pmatrix}
\mathrm{W}(a) \, \mathrm{L}(0,a) =
\begin{pmatrix}
0 & 0
\end{pmatrix}
$
and to substitute (\ref{Lellip}) with the result
$$
\begin{pmatrix}
(z)_4 , & (z)_3
\end{pmatrix} \mathrm{W}(a)
\,\frac{1}{\theta_1(2z)}\,
\mathrm{M}(z\mp a)
%%\begin{pmatrix}
%%(z-a)_3 & -(z+a)_3 \\
%%-(z-a)_4 & (z+a)_4
%%\end{pmatrix}
\begin{pmatrix}
e^{\eta \partial_{z}} & 0 \\
0 & e^{-\eta \partial_{z}}
\end{pmatrix}
\begin{pmatrix}
1 \\ 1
\end{pmatrix}  = 0 \,.
$$
Consequently verifying the system of two relations in (\ref{rec1}) (or in (\ref{rec2}))
one needs to check only the first one. In the following Subsection
proving Baxter equation we will use only (\ref{recEl1}).

We also need a series of relations for $\mathrm{S}^2$ (\ref{ellipS2})
\begin{eqnarray}
\label{recurrence1}
e^{\eta \partial_{z_2}}\, \mathrm{S}^2(a)\, e^{-\eta \partial_{z_2}} & = &
- \mathrm{R}^{-2}(\tau) \, e^{-\frac{\pi i \tau}{2}} \,
\theta^{-1}_4(\mp z_1 - z_2 + a) \,\mathrm{S}^2 (a+\eta)\,, \\[0.2 cm]
\label{recurrence2}
e^{\eta \partial_{z_2}}\, \mathrm{S}^2(a)\, e^{-\eta \partial_{z_2}} & = &
- \mathrm{R}^2(\tau) \, e^{\frac{\pi i \tau}{2}} \,
\theta_4(\mp z_1 + z_2 + a) \,\mathrm{S}^2 (a-\eta)\,, \\[0.2 cm]
\label{recurrence3}
e^{-\eta \partial_{z_2}}\, \mathrm{S}^2(a)\, e^{\eta \partial_{z_2}} & = &
- \mathrm{R}^2(\tau) \, e^{\frac{\pi i \tau}{2}} \,
\theta_4(\mp z_1 - z_2 + a) \,\mathrm{S}^2 (a-\eta)\,.
\end{eqnarray}

All these recurrence relations can be easily
proven taking into account the explicit expression for
intertwining operator $\mathrm{W}$ (\ref{ellipW}).
For example we will prove here
relation (\ref{recbEl}).
Let us consider
\begin{equation} \label{DWPhi}
\left[
  (z+b)_i e^{\eta \partial_z} - (z-b)_i e^{-\eta \partial_z}
\right] \mathrm{W}(b-\eta) \, \Phi(z)\, ,
\end{equation}
where we apply the finite difference operator
to the kernel of the integral operator $\mathrm{W}(b-\eta)$ (\ref{ellipW}) and perform argument
shifts in the elliptic gamma functions
by means of (\ref{GammaShift}) obtaining
$$
\frac{\mathrm{R}(\tau)e^{-\pi i \eta}}{\theta_1(-2b)}
\int_{0}^{1} \mathrm{d} x \, \mu(x)
\frac{e^{-\frac{\pi i}{\eta}(z^2+x^2)}}{\Gamma(-2 b)}\,
\left[(z+b)_i \theta_1(z \mp x-b) -
      (z-b)_i \theta_1(- z \mp x-b)
\right] \Gamma(\mp z \mp x - b) \Phi(x)\,.
$$
Further simplifying the combination of theta functions in the latter formula
using (\ref{th1th3})
$$
(z+b)_i \,\theta_1(z \mp x-b) - (z-b)_i \,\theta_1(- z \mp x-b) =
(x)_i \,\theta_1(2z) \,\theta_1(-2b)\,
$$
we see that (\ref{DWPhi}) is equal to
$\mathrm{R}(\tau) \, e^{-\pi i \eta} \,\theta_1(2z) \,\mathrm{W}(b) \, (z)_3 \, \Phi(z)$
in accordance with (\ref{recbEl}).

In a similar way using the formula for the argument shift in the elliptic gamma function
(\ref{GammaShift}) and the formulae
(\ref{thetatheta}) (or (\ref{th1th3})) relating theta functions with
quasi periods $\tau$ and $\frac{\tau}{2}$ it is not difficult to check the
other  recurrence relations.
Let us emphasize
that we do not need Riemann identities for
theta functions for verifying the recurrence relations. Therefore we do not
need them in proving that $\mathrm{W}$ (\ref{WS}) is indeed the intertwining operator.

%%%%%%%%%%%%%%%%%%%%%%%%%%%%%%%%%%%%%%%%%%%%%%%%%%%%%%%%%%%%%%%%%%%%%%%%%%%%%%%%%%%%%%%
%%%%%%%%%%%%%%%%%%%%%%%%%%%%%%%%%%%%%%%%%%%%%%%%%%%%%%%%%%%%%%%%%%%%%%%%%%%%%%%%%%%%%%%
%%%%%%%%%%%%%%%%%%%%%%%%%%%%%%%%%%%%%%%%%%%%%%%%%%%%%%%%%%%%%%%%%%%%%%%%%%%%%%%%%%%%%%%
\subsection{Factorization of the intertwining operator}
\label{intertwiner}
%%%%%%%%%%%%%%%%%%%%%%%%%%%%%%%%%%%%%%%%%%%%%%%%%%%%%%%%%%%%%%%%%%%%%%%%%%%%%%%%%%%%%%%
Now we digress from the main line of our construction
aiming at Baxter $\mathrm{Q}$-operators for generic values of $2\ell$
and  show an interesting application of the above recurrence relations
 if $2\ell +1 = n \in \NN$.
Since
$\mathrm{W}(0)=\II$ using (\ref{recaEl}), (\ref{recbEl}) we
factorize the intertwining operator $\mathrm{W}(\eta n)$ (\ref{WS})
in the case of representations with parameters $\ell=\frac{n-1}{2}$ and $-\ell-1=-\frac{n+1}{2}$
at $n=1,2,\cdots$
into a product of $n$  simpler
finite-difference operators,
\begin{equation} \label{WFact}
\begin{array}{c}
\mathrm{W}(\eta n) =
%%%% \mathrm{R}^{-n}(\tau)\, e^{\pi i \eta n}\,
\frac{c^n}{(z)^{n}_i} \cdot
\prod_{k=0}^{n-1} \frac{1}{\theta_1(2z)}
\left[(z-\eta k )_i \,e^{\eta \partial_z} -
(z+ \eta k)_i \, e^{-\eta \partial_z}\right]
= \\[0.5 cm] =
%\mathrm{W}(\eta l) =
%%%% \mathrm{R}^{-n}(\tau)\, e^{\pi i \eta n}
c^n
\cdot
\prod_{k=0}^{n-1} \frac{1}{\theta_1(2z)}
\left[(z+\eta n-\eta k )_i \,e^{\eta \partial_z} -
(z- \eta n + \eta k)_i \, e^{-\eta \partial_z}\right]
\cdot \frac{1}{(z)^{n}_i} \;\; ; \;\; i = 3,4\,
\end{array}
\end{equation}
where $c= \mathrm{R}^{-1}(\tau)\, e^{\pi i \eta}$.
For illustration we write this explicitly
for
 $n=1,2$ corresponding to spin $\ell=0$ and $\ell=\frac{1}{2}$ respectively.
\begin{eqnarray*}
\mathrm{W}(\eta) &=& c \cdot
\frac{1}{\theta_1(2z)}
\left[e^{\eta \partial_z} - e^{-\eta \partial_z}\right] \,,
\\
\mathrm{W}(2 \eta) &=& \frac{c^2}{(z)_i} \cdot
\frac{1}{\theta_1(2z)}
\left[e^{\eta \partial_z} - e^{-\eta \partial_z}\right]\cdot
\frac{1}{\theta_1(2z)}
\left[(z-\eta)_i \,e^{\eta \partial_z} -
(z+ \eta)_i \, e^{-\eta \partial_z}\right]\,.
\end{eqnarray*}
%% $$
%% \mathrm{W}(2 \eta) =
%% c^2 \cdot
%% \left[
%% \frac{1}{\theta_1(2 z)\theta_1(2 z+2\eta)}e^{2 \eta \partial_z}
%% - \frac{\theta_1(4 \eta)}{\theta_1(2 \eta)}\frac{1}{\theta_1(2 z-2\eta)\theta_1(2 z+2 \eta)}
%% +\frac{1}{\theta_1(2 z-2\eta)\theta_1(2 z)}e^{-2 \eta \partial_z}
%% \right]
%% $$
Expanding the latter formula we obtain a sum of four finite difference
operators which can be simplified further by means of (\ref{th1th3})
$$
\mathrm{W}(2 \eta) =
\frac{c^2}{\theta_1(2 z-2\eta)\theta_1(2 z)\theta_1(2 z+2\eta)} \cdot
\left[
\theta_1(2 z-2\eta) e^{2 \eta \partial_z}
- \frac{\theta_1(4 \eta)}{\theta_1(2 \eta)} \theta_1(2 z)
+\theta_1(2 z+2\eta)e^{-2 \eta \partial_z}
\right].
$$
Similarly using (\ref{th1th3})
it is rather straightforward
to rewrite (\ref{WFact})
in the form of a sum
$$
\mathrm{W}(\eta n) = c^n \cdot
\sum_{k=0}^{n}
(-)^k
\begin{bmatrix}
n \\ k
\end{bmatrix}
\frac{\theta_1(2 z + 2 \eta n - 4 \eta k)}{\prod_{j = 0}^{n}\theta_1(2 z - 2\eta k + 2 \eta j)}
e^{(n-2k)\eta \partial_z}\;\;;\;\;
\begin{bmatrix}
n \\ k
\end{bmatrix} = \frac{\prod_{j=1}^{n} \theta_1(2 \eta j)}
                     {\prod_{j=1}^{k} \theta_1(2 \eta j) \cdot \prod_{j=1}^{n-k} \theta_1(2 \eta j)}\,.
$$
%%It can be shown by induction that this
%%expression coincides with the one proposed
%%in \cite{} in the form of operator sum.
%Expanding the latter formula we obtain a sum of finite-difference
%operators which coincides with
The latter expression for the
intertwining operator at (half)-integer spin appeared first in \cite{Zab98}
and has been derived directly from the integral operator representation
in \cite{EllDerSp}.

Let us note that this factorized representation (\ref{WFact}) for
$\mathrm{W}(\eta n)$ is a rather special form since
all factors contain the third or the fourth Jacobi theta function
simultaneously, and that
a set of similar ones
follows from the recurrence relations (\ref{recaEl}), (\ref{recbEl}).
Their linear combinations
can be arranged in compact fomulae taking into account (\ref{thetatheta})
\begin{eqnarray*}
\prod_{k=0}^{n-1}
\theta_1 (z \mp a_{k} ) \cdot  \mathrm{W}(\eta n) & = &
c^n \cdot
\prod_{k=0}^{n-1} \frac{1}{\theta_1(2z)}
\left[\theta_1(z-\eta k \mp a_{k}) \,e^{\eta \partial_z} -
\theta_1(z+ \eta k \mp a_{k}) \, e^{-\eta \partial_z}\right]\,,
\\
\mathrm{W}(\eta n) \cdot \prod_{k=0}^{n-1} \theta_1 (z \mp a_k ) & = &
c^n \cdot
\prod_{k=0}^{n-1} \frac{1}{\theta_1(2z)}
\left[\theta_1(z+\eta n-\eta k \mp a_{k}) \,e^{\eta \partial_z} -
\theta_1(z- \eta n + \eta k \mp a_{k}) \, e^{-\eta \partial_z}\right]
\end{eqnarray*}
where $a_0,\cdots,a_{n-1}$ denote arbitrary parameters.

The irreducible representation of the Sklyanin algebra
at (half)-integer spin $\ell=\frac{n-1}{2}$
is $n$-dimensional and it can be realised in the space $\Theta^{+}_{2n-2}$
of even theta functions of order $2n-2$.
By the factorized representation
of $\mathrm{W}(\eta n)$ we see that its action annihilates this
irreducible representation space.
Let us demonstrate this fact by means of the recurrence relations.
The set of $n$ functions
$$
(z)^k_3 \, (z)^{n-1-k}_4 \;\;;\;\; k=0,1,\cdots, n-1
$$
form a basis in the space $\Theta^{+}_{2n-2}$.
Then applying $n-1$ times (\ref{recbEl})
and taking into account that
$$
\mathrm{W}(\eta)\cdot 1 = \frac{c}{\theta_1(2z)}
\left[e^{\eta \partial_z} - e^{-\eta \partial_z}\right]
\cdot 1 = 0
$$
we obtain that  $\mathrm{W}(\eta n) \cdot (z)^k_3 \, (z)^{n-1-k}_4 =0\,$.
\subsection{Elliptic Baxter equation}
\label{ellipBaxEq}
%%%%%%%%%%%%%%%%%%%%%%%%%%%%%%%%%%%%%%%%%%%%%%%%%%%%%%%%%%%%%%%%%%%%%%%%%%%%%%%%%%%%%%%%%%%%%%%%%%%%%%%%%%%%%%%%%%%
Now we have at our disposal all necessary identities to prove the Baxter equation.
The calculation in this Subsection
repeats step by step the one of the Subsection \ref{qBaxEq} devoted to the
$q$-deformed case. Our aim is to
obtain the corresponding local relation
underlying Baxter equation. We compute the matrix elements of
(cf. (\ref{cornerstone}))
\begin{equation} \label{R2sigmaL}
\begin{pmatrix}
1 & 0 \\
(z_1)_4 & -(z_1)_3
\end{pmatrix}
\mathrm{R}^2(u)\,
\sigma_3\,\mathrm{L}(u_1,u_2)
\begin{pmatrix}
(z_2)_3 & 0 \\
(z_2)_4 & -1
\end{pmatrix}
\end{equation}
starting from the defining relation for $\mathrm{R}^2$ (\ref{R2LL}) where
as before $\mathrm{R}^2(u)=\mathrm{R}^2(u_1,u_2|0)$.
Let us remind that in all formulae
we multiply the $\mathrm{L}$-operator (\ref{Lellip})
by the Pauli matrix $\sigma_3$ on the left:
$\mathrm{L}(u)\to \sigma_3\, \mathrm{L}(u)$.
%%%%%%%%%%%%%%%%%%%%%%%%%%%%%%%%%%%%%%%%%%%%%%%%%%%%%%%%%
%%We construct $\mathrm{Q}_2$ (\ref{Q2})
%%out of local operators $\mathrm{R}^2$ and
%%$$
%%\mathrm{R}^2(u_1 , u_2 | v_2) = \mathrm{S}_3 (u_1 - v_2) \, \mathrm{S}_2 (u_2 - v_2) \, \mathrm{S}_3 (u_2 - u_1)
%%$$
%%All steps of the proof below repeat ones in $q$-case.
%%\begin{equation} \label{R2LsigmaL}
%%\mathrm{R}^2_{12}(u_1,u_2|v_2) \, \sigma_3 \, \mathrm{L}_1 (u_1 , u_2) \,\sigma_3\, \mathrm{L}_2 (v_1 , v_2) =
%%\sigma_3\,\mathrm{L}_1 (u_1 , v_2) \,\sigma_3\, \mathrm{L}_2 (v_1 , u_2) \; \mathrm{R}^2_{12}(u_1,u_2|v_2)
%%\end{equation}
%%%%%%%%%%%%%%%%%%%%%%%%%%%%%%%%%%%%%%%%%%%%%%%%%%%%%%%%%%
%%Also
%%our calculation is heavily based on the recurrence relations
We start from \eqref{R2LL} for elliptic deformation case,
use the factorized form of the $\mathrm{L}$-operator
\eqref{Lellip} and take into account the commutativity of $\mathrm{R}^2$ with $z_2$
to obtain
\begin{equation} \label{basicEl}
\mathrm{R}^2(u_1 , u_2 |v_2)\, \sigma_3 \mathrm{L}_1(u_1 , u_2)\, \sigma_3
\mathrm{M}(z_2 \mp v_2)
=
\sigma_3 \mathrm{L}_1 (u_1 , v_2)\,\sigma_3
\mathrm{M}(z_2\mp u_2)
\begin{pmatrix}
e^{\eta \partial_{z_2}} \mathrm{R}^2 e^{-\eta \partial_{z_2}} & 0 \\
0 & e^{-\eta \partial_{z_2}} \mathrm{R}^2 e^{\eta \partial_{z_2}}
\end{pmatrix}
\end{equation}
or in the other form due to (\ref{MN})
\begin{equation} \label{basic2El}
\begin{array}{c}
\mathrm{R}^2(u_1 , u_2 |v_2) \,\sigma_3\, \mathrm{L}_1(u_1 , u_2) \,\sigma_3
= \sigma_3\,\mathrm{L}_1 (u_1 , v_2)\,
\sigma_3\,
\mathrm{M}(z_2\mp u_2) \times
\\[0.3 cm]
\cdot
\begin{pmatrix}
e^{\eta \partial_{z_2}} \mathrm{R}^2(u_1,u_2|v_2) e^{-\eta \partial_{z_2}} & 0 \\
0 & e^{-\eta \partial_{z_2}} \mathrm{R}^2(u_1,u_2|v_2) e^{\eta \partial_{z_2}}
\end{pmatrix}
\mathrm{N}(z_2\mp v_2)\,
\frac{1}{2\theta_1(2 z_2)\,\theta_1(2 v_2)}\,.
\end{array}
\end{equation}
These are the two main relations in the following calculation.

We start with  the matrix element below diagonal of (\ref{R2sigmaL}).
Taking into account (\ref{Lellip}),
$$
\begin{pmatrix}
(z_1)_4 , & -(z_1)_3
\end{pmatrix} \mathrm{L}(u_1,0)
\sim
\begin{pmatrix}
(z_1)_4 , & -(z_1)_3
\end{pmatrix} \mathrm{M}(z_1)=
\begin{pmatrix}
0 , & 0
\end{pmatrix},
$$
and (\ref{basicEl}) at $v_2=0$ we conclude that it is equal to zero in
agreement with general statement (\ref{cornerstone}).
%$$
%\begin{pmatrix}
%(z_1)_4 & -(z_1)_3
%\end{pmatrix}
%\mathrm{R}^2(u_1,u_2|0)\,
%\sigma_3\,\mathrm{L}(u_1,u_2)
%\begin{pmatrix}
%(z_2)_3 \\
%(z_2)_4
%\end{pmatrix}
%$$

Further let us consider the  first diagonal element in (\ref{R2sigmaL}):
$
\begin{pmatrix}
1 , & 0
\end{pmatrix}
\mathrm{R}^2(u) \,\sigma_3 \mathrm{L}(u_1,u_2)
\begin{pmatrix}
(z_2)_3 \\ (z_2)_4
\end{pmatrix}
$.
From (\ref{basicEl}) at $v_2 =0$
we see that it is equal to
$$
\begin{pmatrix}
1 ,& 0
\end{pmatrix}
\mathrm{L}_1(u_1,0)\, \sigma_3\,\mathrm{M}(z_2\mp u_2)
\begin{pmatrix}
1 \\ 0
\end{pmatrix}
\,e^{\eta \partial_{z_2}}\,\mathrm{R}^2(u)\,e^{-\eta \partial_{z_2}}\,.
$$
Then we take into account that (\ref{R2})
%%\begin{equation} \label{R2twistEl}
$
e^{\eta \partial_{z_2}} \,\mathrm{R}^2(u) \, e^{-\eta \partial_{z_2}} =
\mathrm{S}^3 (u_1) \, e^{\eta \partial_{z_2}} \, \mathrm{S}^2( u_2 )
\, e^{-\eta \partial_{z_2}} \, \mathrm{S}^3 ( u_2 - u_1 )
$
%%\end{equation}
and move $\mathrm{S}^3 (u_1)$ to the left by means of intertwining relation (\ref{S3})
$\mathrm{S}^3 (u_1) \,\mathrm{L}(0,u_1) = \mathrm{L}(u_1,0)\,\mathrm{S}^3 (u_1)$
and obtain
$$
\mathrm{S}^3 (u_1)
\begin{pmatrix}
1 ,& 0
\end{pmatrix}
\frac{1}{\theta_1(2z_1)}\,
\mathrm{M}(z_1 \mp u_1)
\begin{pmatrix}
e^{\eta \partial_{z_1}} & 0 \\
0 & e^{-\eta \partial_{z_1}}
\end{pmatrix}
\underline{
\mathrm{N}(z_1) \, \sigma_3
\begin{pmatrix}
(z_2 - u_2)_3 \\ (z_2-u_2)_4
\end{pmatrix} }
e^{\eta \partial_{z_2}} \mathrm{S}^2( u_2 )
e^{-\eta \partial_{z_2}} \mathrm{S}^3 ( u_2 - u_1 ).
$$
The underlined
matrix in the previous formula is equal to
$2 \, \theta_4(z_1+z_2-u_2)\,\theta_4(z_1-z_2+u_2)
\begin{pmatrix}
1 \\1
\end{pmatrix}$ in view of (\ref{M;N}) and (\ref{thetatheta}).
Thus using the recurrence relation (\ref{recurrence1})
we find that the wanted  matrix element is equal to
$$
- 2\,\mathrm{R}^{-2}(\tau)\,e^{-\frac{\pi i \tau}{2}}\,
\underline{
\mathrm{S}^3 (u_1)\, \frac{1}{\theta_1(2z_1)}
\left[ (z_1-u_1)_3 \, e^{\eta \partial_{z_1}} - (z_1+u_1)_3 \, e^{-\eta \partial_{z_1}} \right] }
\mathrm{S}^2( u_2 + \eta)\,\mathrm{S}^3 ( u_2 - u_1 )\,.
$$
Finally  using the recurrence relation (\ref{recEl1}) for the
intertwining operator $\mathrm{S}^3$
in the underlined factor we have (\ref{R2})
$$
- 2\,\mathrm{R}^{-2}(\tau)\,e^{-\pi i \eta-\frac{\pi i \tau}{2}}\,(z_1)_3\,
\mathrm{R}^2_{12}(u+2\eta)\,.
$$

Consider the second diagonal matrix element in (\ref{R2sigmaL}):
$
\begin{pmatrix}
(z_1)_4 , & -(z_1)_3
\end{pmatrix}
\mathrm{R}^2(u) \,\sigma_3 \mathrm{L}(u_1,u_2)
\begin{pmatrix}
0\\ -1
\end{pmatrix}
$.
Due to (\ref{basic2El}) it is equal to
$$
\begin{array}{c}
\begin{pmatrix}
(z_1)_4 , & (z_1)_3
\end{pmatrix}
\mathrm{L}_1(u_1,v_2)\,\sigma_3\,
\mathrm{M}(z_2 \mp u_2)
\times \\ [0.2 cm] \cdot
\begin{pmatrix}
e^{\eta \partial_{z_2}} \, \mathrm{R}^2(u) \, e^{-\eta \partial_{z_2}} & 0 \\
0 & e^{-\eta \partial_{z_2}} \, \mathrm{R}^2(u) \, e^{\eta \partial_{z_2}}
\end{pmatrix}
\begin{pmatrix}
(z_2+v_2)_3 \\
(z_2-v_2)_3
\end{pmatrix}
\frac{1}{2\,\theta_1(2 z_2)\,\theta_1(2 v_2)}
\end{array}
$$
where we have to take carefully the limit $v_2=0$.
To do this we notice that (\ref{M;N}) , (\ref{thetatheta})
$$
\frac{1}{\theta_1(2v_2)}
\begin{pmatrix}
(z_1)_4 , & (z_1)_3
\end{pmatrix} \mathrm{M}(z_1 \mp v_2) =
\frac{2}{\theta_1(2v_2)}
\begin{pmatrix}
\theta_1(2 z_1-v_2) \,\theta_1(v_2) \\
\theta_1(2 z_1+v_2) \,\theta_1(v_2)
\end{pmatrix}
\to \theta_1(2z_1)
\begin{pmatrix}
1 \\ 1
\end{pmatrix} \,\,\,\mbox{at} \,\,v_2 \to 0\,.
$$
Thus the wanted matrix element takes the form (\ref{Lellip})
$$
\frac{1}{2}
\begin{pmatrix}
1 \\ 1
\end{pmatrix}
\begin{pmatrix}
e^{\eta \partial_{z_1}} & 0 \\
0 & e^{-\eta \partial_{z_1}}
\end{pmatrix}
\mathrm{N}(z_1 \mp u_1) \, \mathrm{S}^3(u_1) \, \sigma_3
\, \mathrm{M}(z_2 \mp u_2)
\underline{
\begin{pmatrix}
e^{\eta \partial_{z_2}} \mathrm{S}^2(u_2) \,e^{-\eta \partial_{z_2}} \\
e^{\eta \partial_{z_2}} \mathrm{S}^2(u_2) \,e^{-\eta \partial_{z_2}}
\end{pmatrix} }
\frac{(z_2)_3}{\theta_1(2 z_2)} \,\mathrm{S}^3(u_2-u_1)\,.
$$
Further we take into account the recurrence relations (\ref{recurrence2}),
(\ref{recurrence3}) and (\ref{M;N}), (\ref{thetatheta})
and find that the underlined matrix can be written as follows
$$
-\mathrm{R}^2(\tau)\,e^{\frac{\pi i \tau}{2}}\,\mathrm{S}^2(u_2-\eta)
\, \frac{1}{2} \,\mathrm{N}(z_2 \mp u_2)
\begin{pmatrix}
(z_1)_3 \\
(z_1)_4
\end{pmatrix}.
$$
Since the  matrices $\mathrm{M}$ and $\mathrm{N}$ are
inverse to each other (\ref{MN}):
$\mathrm{M}(z_2 \mp u_2)\,\mathrm{N}(z_2 \mp u_2)
= 2\,\theta_1(2u_2) \theta_1(2z_2) \II$,
the wanted matrix element takes the form
$$
-\frac{1}{2}\,\mathrm{R}^2(\tau)\,e^{\frac{\pi i \tau}{2}}
\theta_1 (2 u_2) (z_2)_3
\begin{pmatrix}
1 \\ 1
\end{pmatrix}
\underline{
\begin{pmatrix}
e^{\eta \partial_{z_1}} & 0 \\
0 & e^{-\eta \partial_{z_1}}
\end{pmatrix}
\mathrm{N}(z_1 \mp u_1)\,
\mathrm{S}^3(u_1)
\begin{pmatrix}
(z_1)_3 \\
-(z_1)_4
\end{pmatrix} }
\mathrm{S}^2(u_2-\eta)\,\mathrm{S}^3(u_2-u_1).
$$
Using the recurrence relation (\ref{recEl1})
we finally obtain the wanted matrix element
$$
-2\,\mathrm{R}^{-1}(\tau)\,e^{\frac{\pi i \tau}{2}+\pi i \eta} \,
\theta_1 (2 u_1) \, \theta_1 (2 u_2) \, (z_2)_3\,
\mathrm{R}^2(u-2\eta)\,,
$$
that completes the calculation of (\ref{R2sigmaL}).

Inserting the  permutation operator $\mathrm{P}_{12}$ and
implementing the similarity transformation in (\ref{R2sigmaL})
we obtain that (cf. (\ref{cornerstone}))
\begin{equation}
\mathrm{Z}_{2}\,
%%\begin{pmatrix}
%%(z_2)^{-1}_3 & 0 \\
%%(z_2)_4 & -(z_2)_3
%%\end{pmatrix}
\mathbb{R}^2_{12}(u)\,
\sigma_3\,\mathrm{L}_1(u_1,u_2)\,
\mathrm{Z}^{-1}_{2}
%%\begin{pmatrix}
%%(z_2)_3 & 0 \\
%%(z_2)_4 & -(z_2)^{-1}_3
%%\end{pmatrix}
=
\begin{pmatrix}
2 \,\kappa^{-1}\mathbb{R}^2_{12}(u+2\eta) & \cdots \\
0 & 2 \,\kappa\,\theta_1(2 u_1)\,\theta_1(2 u_2)\,\mathbb{R}^2_{12}(u-2\eta)
\end{pmatrix}
\end{equation}
where as usual $\mathbb{R}^2_{12}=\mathrm{P}_{12} \,\mathrm{R}^2_{12}$ ,
the  constant is
$\kappa = - \mathrm{R}(\tau) \, e^{\pi i \eta + \frac{\pi i \tau}{2}}$ and
$\mathrm{Z}_2$ stands for
$
\mathrm{Z}_2 =
\begin{pmatrix}
(z_2)^{-1}_3 & 0 \\
(z_2)_4 & -(z_2)_3
\end{pmatrix}
$.
According to the final step explained in Section 2
 we  obtain immediately the Baxter equation (cf. (\ref{BE}))
\begin{equation} \label{ellipBE}
\mathrm{t}(u)\, \mathrm{Q}_2(u) = 2^N \kappa^{-N}\, \mathrm{Q}_2(u+2\eta) +
2^N \kappa^{N}\, \Delta^N(u_1,u_2)\,\mathrm{Q}_2(u-2\eta)
\end{equation}
where we used the  notation
$\Delta(u_1,u_2) = \theta_1(2 u_1)\,\theta_1(2 u_2)$ for a symmetric function.
Here the transfer matrix $\mathrm{t}(u)$ (\ref{tmat}) is constructed out of
$\sigma_3\,\mathrm{L}(u)$ according to our adopted convention.

%%%%%%%%%%%%%%%%%%%%%%%%%%%%%%%%%%%%%%%%%%%%%%%%%%%%%%%%%%%%%%%%%%%%%%%%%%%%%%%%%%%%%%%%%%
%%%%%%%%%%%%%%%%%%%%%%%%%%%%%%%%%%%%%%%%%%%%%%%%%%%%%%%%%%%%%%%%%%%%%%%%%%%%%%%%%%%%%%%%%%
%%%%%%%%%%%%%%%%%%%%%%%%%%%%%%%%%%%%%%%%%%%%%%%%%%%%%%%%%%%%%%%%%%%%%%%%%%%%%%%%%%%%%%%%%%
\subsection{Explicit action of the elliptic $\mathrm{Q}$-operator}
\label{ellipExplQ}
%%%%%%%%%%%%%%%%%%%%%%%%%%%%%%%%%%%%%%%%%%%%%%%%%%%%%%%%%%%%%%%%%%%%%%%%%%%%%%%%%%%%%%%%%%
In the $q$-deformation case we have found in  Subsection \ref{qExplQ}
that the operator $\mathrm{R}^2_{12}$ acts in a  simple way
on a certain function of the variable $z_1$ and of the auxiliary parameter $\lambda$
(\ref{qR2GenFun}).
In Appendix \ref{altern} we also show
that this formula can be obtained at least formally
using the  Coxeter relation (\ref{S3S2S3}) only.
Now we are going to deduce the analogous result in the elliptic case.
Using  the  formulation
with the intertwining operator $\mathrm{W}$ being an integral operator (\ref{ellipW})
we have a solid base to deduce an elliptic analog of the formula (\ref{qR2GenFun})
from Coxeter relation because the latter
is equivalent to the elliptic beta integral evaluation formula.

Thus we start from the Coxeter relation (\ref{S3S2S3})
$
\mathrm{S}^2(u_2-u_1)\,\mathrm{S}^3(u_2)\,\mathrm{S}^2(u_1)\,\mathrm{S}^3(u_1-u_2)=
\mathrm{S}^3(u_1)\,\mathrm{S}^2(u_2)
$
and apply both  sides to $\delta(z_1-z_3)$.
Using the integral operator for $\mathrm{S}^1, \mathrm{S}^3$ (\ref{ellipW})
we assume the position variables  taking real values.

Since (\ref{ellipS2}), (\ref{ellipW})
$$
\mathrm{S}^3(a) \cdot \delta(z_1-z_3)=
\frac{\mathrm{C}\,e^{-\frac{\pi i}{\eta}z_1^2+\frac{\pi i}{\eta}z_3^2}}
     {\Gamma(-2 a)\Gamma(\mp 2 z_3)}\,
\Gamma(\mp z_1 \mp z_3 - a) \;\;;\;\;
%$$
%and
%$$
\mathrm{S}^2(a) \,\delta(z_1-z_3) =
\left.\mathrm{S}^2(a)\right|_{z_1\to z_3} \delta(z_1-z_3)
$$
and $\mathrm{R}^2(u) = \mathrm{S}^2(u_2-u_1)\,\mathrm{S}^3(u_2)\,\mathrm{S}^2(u_1)$ (\ref{R2})
we obtain immediately the wanted local formula
%%$$
%%\begin{array}{c}
%%\mathrm{R}^{2}_{12}(u_1,u_2|0) \cdot
%%e^{-\frac{\pi i}{\eta} z_1^2}
%%\Gamma(\mp z_1 \mp z_3 + u_2 - u_1) =
%%\frac{\Gamma(2 u_2-2 u_1)}{\Gamma(-2 u_1)}\,
%%e^{-\frac{\pi i}{\eta} z_1^2} \,
%%\Gamma(\mp z_1 \mp z_3 - u_1)\,
%%\Gamma(\mp z_2 \mp z_3 + u_2 + \eta + \frac{\tau}{2})
%%\end{array}
%%$$
\begin{equation}
\label{ellipR2GenFun}
\begin{array}{c}
\mathrm{R}^{2}_{12}(u) \cdot
e^{-\frac{\pi i}{\eta} z_1^2}
\Gamma(\mp z_1 \mp z_3 + 2 \eta \ell + \eta) = \\ [0.3 cm] =
c \cdot
e^{-\frac{\pi i}{\eta} z_1^2} \,
\Gamma(\mp z_1 \mp z_3 - \frac{u}{2} + \eta \ell +\eta)\,
\Gamma(\mp z_2 \mp z_3 + \frac{u}{2} + \eta \ell + \eta + \frac{\tau}{2})
\end{array}
\end{equation}
where we take into account definition of spectral parameters (\ref{uEl}) and
denote $c = \frac{\Gamma(4 \eta \ell + 2 \eta)}{\Gamma(-u + 2 \eta \ell + 2 \eta)}$.

We proceed to the action of $\mathrm{Q}_2$. For convenience
 we renormalize  $\mathrm{Q}_2$:
$\mathrm{Q}_2(u) \to c^{-N}\cdot\mathrm{Q}(u)$, such that the Baxter
equation takes the form
$$
\mathrm{t}(u)\, \mathrm{Q}(u) = \Delta_{+}(u)\, \mathrm{Q}(u+2\eta) +
\Delta_{-}(u)\,\mathrm{Q}(u-2\eta)
$$
where
$\Delta_{\pm}(u) = 2^N e^{-\pi i \eta N} e^{\mp N(\pi i u - 2 \pi i \eta \ell + \frac{\pi i \tau}{2})}
\theta_1^N(u \mp 2 \eta \ell)$.
Now we see that  due to (\ref{trA}) the
Baxter operator $\mathrm{Q}(u)$ acts on
the generating function depending on arbitrary
$\lambda_1, \cdots, \lambda_N$ as follows
\begin{gather*} \label{ellipQ}
\mathrm{Q}(u) \cdot \prod_{i=1}^{N}
e^{-\frac{\pi i}{\eta} z_i^2}
\Gamma(\mp z_i \mp \lambda_i + 2 \eta \ell + \eta) =\\
=
\prod_{i=1}^{N}
e^{-\frac{\pi i}{\eta} z_i^2} \,
\Gamma(\mp z_i \mp \lambda_i + \frac{u}{2} + \eta \ell + \eta + \frac{\tau}{2})\,
\Gamma(\mp z_i \mp \lambda_{i+1} - \frac{u}{2} + \eta \ell +\eta)\,.
\end{gather*}
A function similar to $\Gamma(\mp z \mp \lambda + 2 \eta \ell + \eta)$ 
was used in \cite{AA2008} as a generating function for the infinite-dimensional module 
of the Sklyanin algebra.

%%%%%%%%%%%%%%%%%%%%%%%%%%%%%%%%%%%%%%%%%%%%%%%%%%%%%%%%%%%%%%%%%%%%%%%%%%%%%%%%%%%%%%%%
%%%%%%%%%%%%%%%%%%%%%%%%%%%%%%%%%%%%%%%%%%%%%%%%%%%%%%%%%%%%%%%%%%%%%%%%%%%%%%%%%%%%%%%%
%%%%%%%%%%%%%%%%%%%%%%%%%%%%%%%%%%%%%%%%%%%%%%%%%%%%%%%%%%%%%%%%%%%%%%%%%%%%%%%%%%%%%%%%
%%%%%%%%%%%%%%%%%%%%%%%%%%%%%%%%%%%%%%%%%%%%%%%%%%%%%%%%%%%%%%%%%%%%%%%%%%%%%%%%%%%%%%%%
%%%%%%%%%%%%%%%%%%%%%%%%%%%%%%%%%%%%%%%%%%%%%%%%%%%%%%%%%%%%%%%%%%%%%%%%%%%%%%%%%%%%%%%%
%%%%%%%%%%%%%%%%%%%%%%%%%%%%%%%%%%%%%%%%%%%%%%%%%%%%%%%%%%%%%%%%%%%%%%%%%%%%%%%%%%%%%%%%
\section{Summary}

In our approach the case of spin chains where the one-site states form
irreducible infinite-dimensional representations
 is the basic one for the construction. This generic case with $\ell \neq \frac{n}{2},
n=0,1,2,\cdots$ is addressed here  where the $s\ell_2$ symmetry is deformed
in trigonometric or elliptical way.
In the preceding paper concerning the undeformed symmetry also the case
of finite dimensional representations at the chain sites has been treated
by investigating the limits where $2\ell$ approaches nonnegative integer.
The presented here discussion  of the deformed cases does not cover this
finite-dimensional representation case.
The expressions for $\mathrm{R}^1, \mathrm{R}^2$ in terms of $q$-Gamma
functions obtained in the trigonometric case (Appendix B) allow to study
the integer limit in analogy to the undeformed case. Missing the analogous
form for $\mathrm{R}^1, \mathrm{R}^2$ in the elliptic case we face an
obstacle to proceed here by analogy.
Nevertheless in Subsection \ref{intertwiner}
we have given a detailed  discussion of the intertwining operator
of the Sklyanin algebra representations at integer and half-integer spin.
In the trigonometric case we have
realized the representations on the space of polynomials
and the corresponding Yang-Baxter operators have been represented as functions of
Weyl pairs. In the elliptic  we
have formulated operators as integral ones.

The idea of looking for such building blocks and relations in the
undeformed case which have immediate counterparts in the deformed cases
provides the guideline through the increasing complexity. The factors of the
general Yang-Baxter operator related to the elementary transpositions of
representation parameters turn out to be the appropriate building blocks
for this purpose.

In this way we have extended the parallel treatment of quantum spin chains
with symmetry bases on the algebra $s\ell_2$ without deformation,  with
trigonometric  and elliptic deformations beyond the local (one-site)
operators considered earlier to the global chain operators.
Besides of the well-known transfer matrix further generating functions of
conserved charges have been considered, in particular the general transfer
matrix $\mathrm{T}_s(u) $ and two Baxter operators, $\mathrm{Q}_1,
\mathrm{Q}_2 $. The scheme of their construction and the proof of the Baxter
equation, which was applied in the preceding paper to the undeformed case,
has been shown here to work in the deformed cases as well.
The construction results in explicit expressions, in particular the explicit
form of the action of $\mathrm{Q}_2$ on a generating function of spin chain
states has been provided for all cases.

Note that in the elliptic case the $\mathrm{Q}$-operator was constructed by
A.~Zabrodin in \cite{Zab} using a
different method.  It will be very interesting to relate both approaches.

%%%%%%%%%%%%%%%%%%%%%%%%%%%%%%%%%%%%%%%%%%%%%%%%%%%%%%%%%%%%%%%%%%%%%%%%%%%%%%%%%%%%%%%%
%%%%%%%%%%%%%%%%%%%%%%%%%%%%%%%%%%%%%%%%%%%%%%%%%%%%%%%%%%%%%%%%%%%%%%%%%%%%%%%%%%%%%%%%
%%%%%%%%%%%%%%%%%%%%%%%%%%%%%%%%%%%%%%%%%%%%%%%%%%%%%%%%%%%%%%%%%%%%%%%%%%%%%%%%%%%%%%%%
\section*{Acknowledgement}

The work of D.~C. is supported by the Chebyshev Laboratory
(Department of Mathematics and Mechanics, St.-Petersburg State University)
under RF government grant 11.G34.31.0026,
and by Dmitry Zimin's "Dynasty" Foundation. He thanks DAAD for supporting
the visit at Leipzig University. 
The work of S.~D. is supported by RFBR grants 
11-01-00570, 11-01-12037,12-02-91052.
D.~K. is supported by Armenian grant 11-1c028.

%%%%%%%%%%%%%%%%%%%%%%%%%%%%%%%%%%%%%%%%%%%%%%%%%%%%%%%%%%%%%%%%%%%%%%%%%%%%%%%%%%%%%%%%
%%%%%%%%%%%%%%%%%%%%%%%%%%%%%%%%%%%%%%%%%%%%%%%%%%%%%%%%%%%%%%%%%%%%%%%%%%%%%%%%%%%%%%%%
%%%%%%%%%%%%%%%%%%%%%%%%%%%%%%%%%%%%%%%%%%%%%%%%%%%%%%%%%%%%%%%%%%%%%%%%%%%%%%%%%%%%%%%%

\appendix
\renewcommand{\theequation}{\Alph{section}.\arabic{equation}}
\setcounter{table}{0}
\renewcommand{\thetable}{\Alph{table}}

\section*{Appendices}

%%%%%%%%%%%%%%%%%%%%%%%%%%%%%%%%%%%%%%%%%%%%%%%%%%%%%%%%%%%%%%%%%%%%%%%%%%%%%%%%%%%%%%%%
%%%%%%%%%%%%%%%%%%%%%%%%%%%%%%%%%%%%%%%%%%%%%%%%%%%%%%%%%%%%%%%%%%%%%%%%%%%%%%%%%%%%%%%%
%%%%%%%%%%%%%%%%%%%%%%%%%%%%%%%%%%%%%%%%%%%%%%%%%%%%%%%%%%%%%%%%%%%%%%%%%%%%%%%%%%%%%%%%
\section{Special functions for trigonometric and elliptic
deformations}
\setcounter{equation}{0}
\label{SpFun}
%%%%%%%%%%%%%%%%%%%%%%%%%%%%%%%%%%%%%%%%%%%%%%%%%%%%%%%%%%%%%%%%%%%%%%%%%%%%%%%%%%%%%%%%
In this Appendix we collect some useful formulae
concerning
the special functions which we need in our calculations.
The standard $q$-products involving a complex number $q$ ($|q|<1$)
are defined as
\begin{equation} \label{qprod}
(x;q^2) = \prod_{i=0}^{\infty}(1-x \,q^{2 i})\;\;,\;\;
(x;q^2)_k = \prod_{i=0}^{k-1}(1-x \,q^{2 i})\,.
\end{equation}
The $q$-binomial formula at $|z|<1$
\begin{equation} \label{qBinom}
\sum_{n \geq 0} \frac{(a;q^2)_n}{(q^2;q^2)_n} \, z^n = \frac{(a \,z;q^2)}{(z;q^2)}
\end{equation}
produces the expansions
\begin{equation} \label{qexp}
(x;q^2) = \sum_{k\geq 0}\frac{(-)^{k}q^{k(k-1)}}{(q^2;q^2)_k} \, x^k \;\;,\;\;
(x;q^2)^{-1} = \sum_{k\geq 0}\frac{x^k}{(q^2;q^2)_k}\,.
\end{equation}
If operators $\mathbf{u}$ and $\mathbf{v}$ form
a Weyl pair: $\mathbf{u} \, \mathbf{v} = q^2\,\mathbf{v} \, \mathbf{u}$ then
\cite{Dilog,Vol,Kir}
\begin{equation} \label{Schutzenberger}
(\mathbf{u};q^2) \, (\mathbf{v};q^2) = (\mathbf{u}+\mathbf{v};q^2)\,,
\end{equation}
\begin{equation} \label{pentagon}
(\mathbf{v};q^2) \, (\mathbf{u};q^2) = (\mathbf{u};q^2)\,(-\mathbf{v\,u};q^2)\,
(\mathbf{v};q^2)\,.
\end{equation}

We use standard definitions of Jacobi theta functions
$\theta_n(z|\tau)$ ($n=1,\cdots,4$) (see e.g. \cite{Erdelyi,aar})
and  the shorthand notations
$\theta_1(z) = \theta_1(z|\tau)$ ,
$\theta_4(z) = \theta_4(z|\tau)$
for theta functions with quasi period $\tau$
and the shorthand notations
$(z)_3 = \theta_3(z|\frac{\tau}{2})$ , $(z)_4 = \theta_4(z|\frac{\tau}{2})$
for theta functions with quasi period $\frac{\tau}{2}$.
Recall that the first theta function is odd $\theta_1(z) = - \theta_1(z)$
and the other three Jacobi theta functions
are even. All Jacobi theta functions are connected
by argument shifts. For example
\begin{equation*}
%% \label{thetaShift}
\begin{array}{c}
\theta_1(z+\frac{\tau}{2}) = i e^{-\pi i z} e^{-\frac{\pi i \tau}{4}} \theta_4(z)\,.
\end{array}
\end{equation*}
Theta functions with quasi periods $\tau$ and $\frac{\tau}{2}$
are related by the bilinear relations
\begin{equation} \label{thetatheta}
\begin{array}{c}
2 \,\theta_1(x\mp y) =
(x)_4 (y)_3 - (y)_4 (x)_3\,, \\[0.3 cm]
2 \,\theta_4(x\mp y) =
(x)_4 (y)_3 + (y)_4 (x)_3\,,
\end{array}
\end{equation}
where we adopt the notation $\theta_{\alpha}(x \mp y)=\theta_{\alpha}(x + y)\, \theta_{\alpha}(x - y)$.
As an immediate consequence of (\ref{thetatheta}) we obtain the formulae
\begin{equation}
\label{th1th3}
(y)_i \, \theta_1(x \mp z) - (x)_i \, \theta_1(y \mp z)  = (z)_i \, \theta_1(x \mp y)
\;\;\;\; \text{where}\;\; i = 3,4\,.
%%\,, \\ [0.3 cm]
%% \label{th4th3}
%% (y)_i \, \theta_4(x \mp z) - (x)_i \, \theta_4(y \mp z) &=& (-)^{\delta_{i,4}} \, (z)_i \, \theta_1(x \mp y)
%% \;\;\;\; \text{where}\;\; i = 3,4\,.
\end{equation}
The elliptic gamma function is defined by the double-infinite product \cite{R,FV,spi:essays}
\begin{equation} \label{elliptG}
\Gamma(z|\tau,2\eta) = \prod_{n,m=0}^{\infty}
\frac{1-e^{2 \pi i (\tau(n+1)+2 \eta (m+1)-z)}}{1-e^{2 \pi i (\tau n + 2 \eta m +
z)}}
\end{equation}
for $\mathrm{Im}\, \tau >0 \,,\mathrm{Im}\, \eta >0$.
In our study we only need its transformation property
under the shift of the argument
\begin{equation} \label{GammaShift}
\Gamma(z+2 \eta) = \mathrm{R}(\tau) \, e^{\pi i z} \, \theta_1 (z) \,
\Gamma(z),
\end{equation}
where $\mathrm{R}(\tau) = - i \,e^{-\frac{\pi i \tau}{4}} (e^{2 \pi i \tau} ; e^{2 \pi i \tau})^{-1}$.
%%%%%%%%%%%%%%%%%%%%%%%%%%%%%%%%%%%%%%%%%%%%%%%%%%%%%%%%%%%%%%%%%%%%%%%%%%%%%%%%%%%%%%%%
%%%%%%%%%%%%%%%%%%%%%%%%%%%%%%%%%%%%%%%%%%%%%%%%%%%%%%%%%%%%%%%%%%%%%%%%%%%%%%%%%%%%%%%%
%%%%%%%%%%%%%%%%%%%%%%%%%%%%%%%%%%%%%%%%%%%%%%%%%%%%%%%%%%%%%%%%%%%%%%%%%%%%%%%%%%%%%%%%
\section{The operators $\mathrm{R}^1$ and $\mathrm{R}^2$ in the  trigonometric
case}
\setcounter{equation}{0}
\label{R1R2}
%%%%%%%%%%%%%%%%%%%%%%%%%%%%%%%%%%%%%%%%%%%%%%%%%%%%%%%%%%%%%%%%%%%%%%%%%%%%%%%%%%%%%%%%
Here we shall establish relations
between several explicit expressions for operators
$\mathrm{R}^1$ and $\mathrm{R}^2$ which respect relations
(\ref{R1LL}) and (\ref{R2LL}), respectively.
In (\ref{R1}) and (\ref{R2}) we have cited several expressions for them.
Let us consider $\mathrm{R}^2$ and take into account (\ref{qW}), (\ref{qS2})
\begin{equation} \label{S2S3S2}
\mathrm{S}^2 (a) \, \mathrm{S}^3 (a+b) \, \mathrm{S}^2 (b) =
c \cdot \frac{(\mathbf{u}_3;q^2)}{(\mathbf{u}_1;q^2)} \, (\mathbf{v}_1;q^2) \,
q^{-(a+b) z_1 \partial_{z_1}} \,
(\mathbf{v}_2;q^2)^{-1} \, \frac{(\mathbf{u}_2;q^2)}{(\mathbf{u}_4;q^2)}
\end{equation}
where $c$ is the  constant $c = q^{\frac{a^2}{2}-\frac{b^2}{2}}$ and
$\mathbf{u}_i$ and $\mathbf{v}_j$ form Weyl pairs:
$\mathbf{u}_i \, \mathbf{v}_j = q^2 \, \mathbf{v}_j \, \mathbf{u}_i$\,,
$$
\mathbf{u}_1 = \frac{z_2}{z_1}q^{1+a} \;;\; \mathbf{u}_2 = \frac{z_2}{z_1}q^{1-b}\;;\;
\mathbf{u}_3 = \frac{z_2}{z_1}q^{1-a} \;;\; \mathbf{u}_4 = \frac{z_2}{z_1}q^{1+b}\;;\;
\mathbf{v}_1 = q^{2 z_1 \partial_{z_1}+2-2a} \;;\; \mathbf{v}_2 = q^{2 z_1 \partial_{z_1}+2+2b}\,.
$$
We are going to rewrite (\ref{S2S3S2}) in several equivalent forms.
By means of the pentagon relation (\ref{pentagon})
$(\mathbf{u};q^2)^{-1} (\mathbf{v};q^2) = (\mathbf{v}- \mathbf{v} \, \mathbf{u} \,;q^2) (\mathbf{u};q^2)^{-1}$
we have
$$
(\ref{S2S3S2}) = c \cdot (\mathbf{u}_3;q^2) \, (\mathbf{v}_1- \mathbf{v}_1 \, \mathbf{u}_1 \,;q^2) \,
\underline{(\mathbf{u}_1;q^2)^{-1} \, q^{-(a+b) z_1 \partial_{z_1}} \, (\mathbf{u}_2;q^2) }
\, (\mathbf{v}_2 - \mathbf{v}_2 \, \mathbf{u}_2 \,;q^2)^{-1} \, (\mathbf{u}_4;q^2)^{-1}\,.
$$
Further we note that underlined expression in the previous formula
is equal to $q^{-(a+b) z_1 \partial_{z_1}}$ and use
Sch\"{u}tzenberger formula (\ref{Schutzenberger})
to rewrite it as follows
\begin{equation} \label{2Fact}
(\ref{S2S3S2}) = c \cdot (\mathbf{u}_3 + \mathbf{v}_1- \mathbf{v}_1 \, \mathbf{u}_1 \,;q^2) \,
\, q^{-(a+b) z_1 \partial_{z_1}} \,
\, (\mathbf{u}_4 + \mathbf{v}_2 - \mathbf{v}_2 \, \mathbf{u}_2 \,;q^2)^{-1}\,.
\end{equation}
Thus we have transformed the original expression (\ref{S2S3S2})
containing $6$ $q$-exponents to the expression with $2$
$q$-exponents.

Further we shall obtain another expression with $4$ $q$-exponents. We
use the pentagon formula (\ref{pentagon})
$(\mathbf{v} - \mathbf{v} \, \mathbf{u} \,;q^2) = (\mathbf{u};q^2)^{-1} \, (\mathbf{v};q^2) \, (\mathbf{u};q^2)$
to transform the previous formula as follows
%%$$
%%c \cdot (\mathbf{u}_1 - \mathbf{v}^{-1}_1 \, \mathbf{u}_3 \,;q^2)^{-1} \, (\mathbf{v_1};q^2) \,
%%\underline{
%%(\mathbf{u}_1 - \mathbf{v}^{-1}_1 \, \mathbf{u}_3 \,;q^2) \, q^{-(a+b) z_1 \partial_{z_1}} \,
%%(\mathbf{u}_2 - \mathbf{v}^{-1}_2 \, \mathbf{u}_4 \,;q^2) } \,
%%(\mathbf{v_2};q^2)^{-1} \, (\mathbf{u}_2- \mathbf{v}^{-1}_2 \, \mathbf{u}_4 \,;q^2) =
%%$$
$$
c \cdot (\mathbf{u}_1 - \mathbf{v}^{-1}_1 \mathbf{u}_3 ;q^2)^{-1} (\mathbf{v_1};q^2)
\underline{
(\mathbf{u}_1 - \mathbf{v}^{-1}_1 \mathbf{u}_3;q^2) q^{-(a+b) z_1 \partial_{z_1}}
(\mathbf{u}_2 - \mathbf{v}^{-1}_2 \mathbf{u}_4 ;q^2) }
(\mathbf{v_2};q^2)^{-1} (\mathbf{u}_2- \mathbf{v}^{-1}_2 \mathbf{u}_4 ;q^2)
$$
and take into account that the underlined expression is equal to
$q^{-(a+b) z_1 \partial_{z_1}}$ to obtain
\begin{equation} \label{4Fact}
c \cdot (\mathbf{u}_1 - \mathbf{v}^{-1}_1 \, \mathbf{u}_3 \,;q^2)^{-1} \, (\mathbf{v_1};q^2) \,
q^{-(a+b) z_1 \partial_{z_1}} \,
(\mathbf{v_2};q^2)^{-1} \, (\mathbf{u}_2- \mathbf{v}^{-1}_2 \, \mathbf{u}_4 \,;q^2)\,.
\end{equation}
As the result  we have found that (\ref{S2S3S2})
can be rewritten
in two equivalent forms (\ref{2Fact}) and (\ref{4Fact}) which appeared in \cite{DKK}
\begin{itemize}
\item
\begin{equation} \label{2Fact'}
\mathrm{S}^2 (a) \, \mathrm{S}^3 (a+b) \, \mathrm{S}^2 (b) =
c\cdot \left( \mathbf{U}(a) ;q^2\right)
q^{-(a+b) z_1 \partial_{z_1}}
\left( \mathbf{U}(-b) ;q^2\right)^{-1}\,,
\end{equation}
where
$$
\mathbf{U}(a) = \frac{z_2}{z_1} q^{1-a} +
q^{2z_1 \partial_{z_1}+2-2a} - \frac{z_2}{z_1} q^{2 z_1 \partial_{z_1}+1-a}\,.
$$
\item
\begin{equation} \label{4Fact'}
\mathrm{S}^2 (a) \, \mathrm{S}^3 (a+b) \, \mathrm{S}^2 (b) =
c\cdot
\left(q^{1+a} \,\overline{\mathbf{u}}\, ; q^2\right)^{-1}
q^{-(a+b) z_1 \partial_{z_1}}
\frac{\left( q^{2 - 2 a} \,\overline{\mathbf{v}}\, ; q^2 \right)}
     {\left( q^{2 + 2 b} \,\overline{\mathbf{v}}\, ; q^2 \right)}
\left(q^{1-b} \, \overline{\mathbf{u}} \, ; q^2\right)\,,
\end{equation}
where
$$
\overline{\mathbf{u}} = \frac{z_2}{z_1}(1-q^{-2z_1\partial_{z_1}}) \;\;;\;\;
\overline{\mathbf{v}} = q^{2 z_1 \partial_{z_1}}\,.
$$
\end{itemize}

%\vspace{0.5 cm}
Now we turn to the operator $\mathrm{R}^1$ (\ref{R1}).
At first let us mention that we cannot rewrite it in the form
like (\ref{S2S3S2}), because the  multipliers $z^a_1$ and $z^b_1$ from
$\mathrm{S}^2(a)$ and $\mathrm{S}^2(b)$ (\ref{qS2})
do not compensate the multiplier $z^{-a-b}_2$ from $\mathrm{S}^1(a+b)$ (\ref{qW}).
To overcome this difficulty we remind
that the defining relation (\ref{S2LL}) does not fix uniquely the operator
$\mathrm{S}^2$:
we can multiply $\mathrm{S}^2$ by an arbitrary function $\varphi$
which respects (\ref{period}).
Now we choose $\mathrm{S}'^2$ (\ref{qS2'}) instead of $\mathrm{S}^2$ (\ref{qS2}).
Since $\mathrm{S}^2 \to \mathrm{S}'^2$
and $\mathrm{S}^3 \to \mathrm{S}^1$
amount to the the change $z_1 \leftrightarrow z_2$
the above calculation for $\mathrm{R}^2$ is suitable as well for
$\mathrm{R}^1$ after the indicated change.

Let us remark that in the original form of
operators $\mathrm{S}^1,\mathrm{S}^2,\mathrm{S}^3$ they respect Coxeter relations
(\ref{S1S2S1}), (\ref{S3S2S3})
and are sufficient to build the general $\mathrm{R}$-operator.
However
 (\ref{4Fact'}) is crucial when we restrict
the general $\mathrm{R}$-operator to the invariant subspace in order
to reproduce the $\mathrm{L}$-operator. Thus if we dealt with
the original set of elementary operators $\mathrm{S}^1,\mathrm{S}^2,\mathrm{S}^3$
only we would not be able to reproduce the standard
$\mathrm{L}$-operator (\ref{Lq}).
%%%%%%%%%%%%%%%%%%%%%%%%%%%%%%%%%%%%%%%%%%%%%%%%%%%%%%%%%%%%%%%%%%%%%%%%%%%%%%%%%%%%%%%%
%%%%%%%%%%%%%%%%%%%%%%%%%%%%%%%%%%%%%%%%%%%%%%%%%%%%%%%%%%%%%%%%%%%%%%%%%%%%%%%%%%%%%%%%
%%%%%%%%%%%%%%%%%%%%%%%%%%%%%%%%%%%%%%%%%%%%%%%%%%%%%%%%%%%%%%%%%%%%%%%%%%%%%%%%%%%%%%%%
\section{From pentagon to Coxeter relations}
\setcounter{equation}{0}
\label{qCoxeter}
%%%%%%%%%%%%%%%%%%%%%%%%%%%%%%%%%%%%%%%%%%%%%%%%%%%%%%%%%%%%%%%%%%%%%%%%%%%%%%%%%
Here we  shall prove the Coxeter relations (\ref{S1S2S1}), (\ref{S3S2S3})
for the elementary intertwining operators $\mathrm{S}^1,\,\mathrm{S}^2,\,\mathrm{S}^3$
in the case of $q$-deformation
using only the  pentagon relation (\ref{pentagon})
for a Weyl pair
$$
\mathbf{u} = \frac{z_2}{z_1}q \;\;;\;\; \mathbf{v} = q^{2 z_1 \partial_{z_1}+2} \;\; ; \;\;
\mathbf{u} \, \mathbf{v} = q^2 \, \mathbf{v} \, \mathbf{u}\,.
$$
We start with the right hand side of (\ref{S3S2S3}),
take into account (\ref{qS2}), (\ref{qW}) and
$c=q^{\frac{a^2}{2}-\frac{b^2}{2}}$,
\begin{equation} \label{s2s3s2}
\mathrm{S}^2(b) \, \mathrm{S}^3(a+b) \, \mathrm{S}^2(a) = c \cdot
\frac{(\mathbf{u}q^{-b};q^2)}{(\mathbf{u} q^{b};q^2)} \,
(\mathbf{v} q^{2a};q^2)^{-1} \cdot q^{-(a+b)z_1 \partial_{z_1}} \cdot (\mathbf{v} q^{-2b};q^2)\,
\frac{(\mathbf{u} q^{-a};q^2)}{(\mathbf{u} q^{a};q^2)}\,.
\end{equation}
Then we apply twice the
pentagon relation (\ref{pentagon}) in the left hand side of the previous
formula
\begin{equation} \label{Ls2s3s2}
\begin{array}{c}
(\mathbf{u}q^{-b};q^2)\,
\underline{(\mathbf{u} q^{b};q^2)^{-1} (\mathbf{v} q^{2a};q^2)^{-1}}=
\underline{(\mathbf{u}q^{-b};q^2) \, (\mathbf{v}q^{2a};q^2)^{-1}}
(-\mathbf{v\,u}\, q^{2a+b} ;q^2)^{-1}
(\mathbf{u}q^{b};q^2)^{-1}
= \\ [0.3 cm] =
(\mathbf{v}q^{2a};q^2)^{-1}
(\mathbf{u}q^{-b};q^2)
\underline{\underline{(-\mathbf{v \,u}\,q^{2a-b};q^2)}}
(-\mathbf{v\,u}\, q^{2a+b} ;q^2)^{-1}
\underline{\underline{(\mathbf{u}q^{b};q^2)^{-1}}}
\end{array}
\end{equation}
and also in the right hand side
\begin{equation} \label{Rs2s3s2}
\begin{array}{c}
\underline{(\mathbf{v} q^{-2b};q^2)\,
(\mathbf{u} q^{-a};q^2)}\,(\mathbf{u} q^{a};q^2)^{-1} =
(\mathbf{u} q^{-a};q^2)\,(-\mathbf{v\,u}\, q^{-a-2b};q^2)\,
\underline{(\mathbf{v} q^{-2b};q^2)\,(\mathbf{u} q^{a};q^2)^{-1}}
= \\ [0.3 cm] =
\underline{\underline{(\mathbf{u} q^{-a};q^2)}}
\,(-\mathbf{v\,u}\, q^{-a-2b};q^2)\,
\underline{\underline{(-\mathbf{v\,u} \,q^{a-2b};q^2)^{-1}}}
(\mathbf{u} q^{a};q^2)^{-1} (\mathbf{v} q^{-2b};q^2)\,.
\end{array}
\end{equation}
Then returning  to (\ref{s2s3s2}) we note that the
double underlined factors in (\ref{Ls2s3s2}) and (\ref{Rs2s3s2})
cancel each other. Thus we rewrite (\ref{s2s3s2}) as
$$
c\cdot (\mathbf{v}q^{2a};q^2)^{-1}
\underline{(\mathbf{u}q^{-b};q^2)\,
(-\mathbf{v\,u}\, q^{-b};q^2)}
\cdot q^{-(a+b)z_1 \partial_{z_1}} \cdot
\underline{(-\mathbf{v\,u}\, q^{a};q^2)^{-1}
(\mathbf{u}q^{a};q^2)^{-1}}
(\mathbf{v}q^{-2b};q^2)=
$$
and apply the pentagon relation (\ref{pentagon})
to obtain that the left hand side of (\ref{S3S2S3}) is
$$
= c\cdot
(\mathbf{v}q^{2a};q^2)^{-1}
(\mathbf{v};q^2)\,
(\mathbf{u}q^{-b};q^2)
\cdot q^{-(a+b)z_1 \partial_{z_1}} \cdot
(\mathbf{u}q^{a};q^2)^{-1}
(\mathbf{v};q^2)^{-1}
(\mathbf{v}q^{-2b};q^2)=
\mathrm{S}^3(a) \, \mathrm{S}^2(a+b) \, \mathrm{S}^3(b)\,.
$$
Similarly we prove the second
Coxeter relation (\ref{S1S2S1})
using a Weyl pair
$$
\overline{\mathbf{u}} = q^{2 z_2 \partial_{z_2}+2} \;\;;\;\;
\overline{\mathbf{v}} = \frac{z_2}{z_1}q \;\; ; \;\;
\overline{\mathbf{u}} \, \overline{\mathbf{v}} =
q^2 \, \overline{\mathbf{v}} \, \overline{\mathbf{u}}\,.
$$
In view of (\ref{qW}), (\ref{qS2})
the left hand side of (\ref{S1S2S1})
 has the explicit form ($c=q^{\frac{(a+b)^2}{2}}$)
\begin{equation} \label{s1s2s1}
\mathrm{S}^1(a) \mathrm{S}^2(a+b) \mathrm{S}^1(b) =
c z^{a+b}_1\cdot
\frac{(\overline{\mathbf{u}};q^2)}{(\overline{\mathbf{u}} q^{2a};q^2)} \,
(\overline{\mathbf{v}} q^{b};q^2)^{-1} \cdot \frac{1}{z^{a+b}_2}q^{-(a+b)z_2 \partial_{z_2}} \cdot
(\overline{\mathbf{v}} q^{-a};q^2)\,
\frac{(\overline{\mathbf{u}} q^{-2b};q^2)}{(\overline{\mathbf{u}};q^2)}\,.
\end{equation}
As before applying the pentagon relation twice in the left hand side and in the
right hand side of the latter formula we have
$$
\begin{array}{ccc}
(\overline{\mathbf{u}};q^2)\,(\overline{\mathbf{u}} q^{2a};q^2)^{-1}
(\overline{\mathbf{v}} q^{b};q^2)^{-1} &=&
(\overline{\mathbf{v}} q^{b};q^2)^{-1}
(\overline{\mathbf{u}} ;q^2)\,
\underline{\underline{(-\overline{\mathbf{v}}\,\overline{\mathbf{u}} \, q^{b};q^2)}}\,
(-\overline{\mathbf{v}}\,\overline{\mathbf{u}} \, q^{2a+b};q^2)^{-1}
\underline{\underline{(\overline{\mathbf{u}} q^{2a};q^2)^{-1}}}
\\ [0.3 cm]
(\overline{\mathbf{v}} q^{-a};q^2)\,
(\overline{\mathbf{u}} q^{-2b};q^2)\,(\overline{\mathbf{u}};q^2)^{-1} &=&
\underline{\underline{(\overline{\mathbf{u}} q^{-2b};q^2)}}\,
(-\overline{\mathbf{v}}\,\overline{\mathbf{u}} \, q^{-a-2b};q^2)\,
\underline{\underline{(-\overline{\mathbf{v}}\,\overline{\mathbf{u}} \, q^{-a};q^2)^{-1}}}
(\overline{\mathbf{u}} ;q^2)^{-1}
(\overline{\mathbf{v}} q^{-a};q^2)\,,
\end{array}
$$
where double underlined factors cancel each other when
we substitute both above formulae in (\ref{s1s2s1}). Thus
(\ref{s1s2s1}) acquires the form
$$
c z^{a+b}_1\cdot
(\overline{\mathbf{v}} q^{b};q^2)^{-1}
\underline{(\overline{\mathbf{u}} ;q^2)\,
(-\overline{\mathbf{v}}\,\overline{\mathbf{u}} \, q^{-b};q^2)}
\cdot \frac{1}{z^{a+b}_2}q^{-(a+b)z_2 \partial_{z_2}} \cdot
\underline{(-\overline{\mathbf{v}}\,\overline{\mathbf{u}} \, q^{a};q^2)^{-1}
(\overline{\mathbf{u}} ;q^2)^{-1}}
(\overline{\mathbf{v}} q^{-a};q^2)\,,
$$
and using the pentagon relation (\ref{pentagon})
two more times we  have finally
$$
c z^{a+b}_1
(\overline{\mathbf{v}} q^{b};q^2)^{-1}
(\overline{\mathbf{v}} q^{-b};q^2)
(\overline{\mathbf{u}} ;q^2)
\frac{1}{z^{a+b}_2}q^{-(a+b)z_2 \partial_{z_2}}
(\overline{\mathbf{u}} ;q^2)^{-1}
(\overline{\mathbf{v}} q^{a};q^2)^{-1}
(\overline{\mathbf{v}} q^{-a};q^2) =
\mathrm{S}^2(b)\mathrm{S}^1(a+b)\mathrm{S}^2(a).
$$
%%%%%%%%%%%%%%%%%%%%%%%%%%%%%%%%%%%%%%%%%%%%%%%%%%%%%%%%%%%%%%%%%%%%%%%%%%%%%%%%%%%%%%%%
%%%%%%%%%%%%%%%%%%%%%%%%%%%%%%%%%%%%%%%%%%%%%%%%%%%%%%%%%%%%%%%%%%%%%%%%%%%%%%%%%%%%%%%%
%%%%%%%%%%%%%%%%%%%%%%%%%%%%%%%%%%%%%%%%%%%%%%%%%%%%%%%%%%%%%%%%%%%%%%%%%%%%%%%%%%%%%%%%
\section{$\mathrm{L}$-operator recovered from the general $\mathrm{R}$-operator
in the trigonometric case }
\setcounter{equation}{0}
\label{R->L}
%%%%%%%%%%%%%%%%%%%%%%%%%%%%%%%%%%%%%%%%%%%%%%%%%%%%%%%%%%%%%%%%%%%%%%%%%%%%%%%%%%%%%%%%
The operator $\mathrm{R}_{12}(u|\ell,s)$ acts in the tensor product
$\mathbb{V}_{\ell} \, \otimes \mathbb{V}_{s} \approx \mathbb{C}[z_1]\otimes\mathbb{C}[z_2]$ of
two infinite-dimensional spaces. At (half)-integer $s$ the space
$\mathbb{V}_{s}$ contains an invariant finite-dimensional subspace $\mathbb{C}^{2s+1}$.
Now we  choose $s=\frac{1}{2}$ and  restrict the general $\mathrm{R}$-operator
to the subspace $\mathbb{V}_{\ell}\otimes\mathbb{C}^2$ of functions of the form
\begin{equation} \label{Fspace}
\Psi(z_1,z_2) = \phi(z_1)+\psi(z_1) \,z_2\,,
\end{equation}
where $\phi$ and $\psi$ are polynomials.
For technical reasons it will be convenient to start with
$2s = 1 - \varepsilon$ and to consider
the limit $\varepsilon \to 0$ in the second factorized form (\ref{R}) of the $\mathrm{R}$-operator
\begin{equation} \label{R2R1}
\mathrm{R}(u_1,u_2|v_1,v_2) = \mathrm{R}^2(v_1,u_2|v_2) \,
\mathrm{R}^1(u_1|v_1,v_2)\,,
\end{equation}
where
\begin{equation} \label{parameter}
u_1 = u -\ell - 1 \; ; \; u_2 = u + \ell \;\;;\;\;
v_1 = -1 - \frac{1}{2} + \frac{\varepsilon}{2} \; ; \; v_2 =
\frac{1}{2} - \frac{\varepsilon}{2}\,.
\end{equation}
The operators $\mathrm{R}^1$ and $\mathrm{R}^2$ are taken
in the form (\ref{4Fact'}) with four $q$-exponential factors.
We start with
\begin{equation} \label{R1uv}
\mathrm{R}^1(u_1|v_1,v_2) =
\left(q^{3-\varepsilon} \,\mathbf{u}\, ; q^2\right)^{-1}
q^{(v_1-u_1) z_2 \partial_{z_2}}
\frac{\left( q^{- 2 + 2 \varepsilon} \,\mathbf{v}\, ; q^2 \right)}
     {\left( q^{ 2 u_1 - 2 v_2 +2} \,\mathbf{v}\, ; q^2 \right)}
\left(q^{v_2-u_1+1} \, \mathbf{u} \, ; q^2\right),
\end{equation}
where
\begin{equation}
\mathbf{u} = \frac{z_1}{z_2}(1-q^{-2z_2\partial_{z_2}}) \;\;;\;\;
\mathbf{v} = q^{2 z_2 \partial_{z_2}}\,,
\end{equation}
and act on the function (\ref{Fspace}). It is clear that due
to the special dependence of the operator $\mathrm{R}^1$ on the variable $z_1$
for our purposes it will be sufficient to apply the operator to the monomials $1$ and $z_2$.
Further we will need the formulae (\ref{qexp})
\begin{equation}
\label{qexpShort}
(x;q^2) = 1 - \frac{x}{1-q^2} + O(x^2) \;\;;\;\;
(x;q^2)^{-1} = 1 + \frac{x}{1-q^2} + O(x^2)\,.
\end{equation}
Thus due to $\mathbf{u}\cdot 1 = 0$ we have
$
\mathrm{R}^1(u_1|v_1,v_2) \cdot 1 =
\frac{\left( q^{- 2 + 2 \varepsilon} ; q^2 \right)}
     {\left( q^{ 2 u_1 - 2 v_2 +2} ; q^2 \right)}
$.
Similarly due to $\mathbf{u}^2 \cdot z_2 =0$ we have
$$
\left(q^{v_2-u_1+1} \, \mathbf{u} \, ; q^2\right) \cdot z_2 = z_2 + q^{v_2-u_1-1} z_1 \;\;;\;\;
\left(q^{3-\varepsilon} \,\mathbf{u}\, ; q^2\right)^{-1} \cdot z_2 = z_2 - q^{1-\varepsilon} z_1
$$
and consequently after some trivial algebra we obtain
$$
\mathrm{R}^1(u_1|v_1,v_2) \cdot z_2 =
\frac{(q^{2\varepsilon};q^2)}{(q^{2 u_1-2v_2+2};q^2)}
\left[ q^{u_1-v_2+1}(1-q^{2v_2-2u_1-4})
\cdot z_1 + q^{v_1-u_1}(1-q^{2u_1-2v_2+2})\cdot z_2  \right] +
O(\varepsilon^2).
$$
Thus we see that $\mathrm{R}^1 \cdot 1 = O(\varepsilon)$ and
$\mathrm{R}^1 \cdot z_2= O(\varepsilon)$. It is due to the factor
$\left( q^{- 2 + 2 \varepsilon} \,\mathbf{v}\, ; q^2 \right)$ in (\ref{R1uv}).
Consequently, to obtain
$\mathrm{R} \cdot \Psi(z_1,z_2)$ at $\varepsilon =0$
we only need to extract the simple poles contributions
from $\mathrm{R}^2$.

Then we consider the second factor in (\ref{R2R1}):
\begin{equation} \label{R2uv}
\mathrm{R}^2(v_1,u_2|v_2) =
\left(q^{u_2-v_1+1} \,\overline{\mathbf{u}}\, ; q^2\right)^{-1}
q^{(v_2-u_2) z_1 \partial_{z_1}}
\frac{\left( q^{ 2 v_1 - 2 u_2 +2} \,\overline{\mathbf{v}}\, ; q^2 \right)}
     {\left( q^{- 2 + 2 \varepsilon} \,\overline{\mathbf{v}}\, ; q^2 \right)}
\left(q^{3 - \varepsilon} \, \overline{\mathbf{u}} \, ; q^2\right),
\end{equation}
where
\begin{equation}
\overline{\mathbf{u}} = \frac{z_2}{z_1}(1-q^{-2z_1\partial_{z_1}}) \;\;;\;\;
\overline{\mathbf{v}} = q^{2 z_1 \partial_{z_1}}\,.
\end{equation}
In the previous formula only the factor
$\left( q^{- 2 + 2 \varepsilon} \,\mathbf{v}\, ; q^2 \right)^{-1}$
can produce poles. As far as we are interested only in singular
contributions
from $\mathrm{R}^2$ we can choose $\varepsilon =0$ in the other
factors of $\mathrm{R}^2$.
The operator $\mathrm{R}^2$ acts trivially on the variable $z_2$
but it acts in a rather nontrivial way on the functions $\phi(z_1)$.
Further we apply it to monomials $z_1^m$.
Due to
$$
\overline{\mathbf{u}}^k \cdot z_1^m =
(-)^k q^{k(k-1)-2mk}\frac{(q^2;q^2)_m}{(q^2;q^2)_{m-k}} \, z_1^{m-k} z_2^k
$$
and (\ref{qexp}) we have
$$
\left(q^{3} \, \overline{\mathbf{u}} \, ; q^2\right) \cdot z_1^m =
\sum_{k=0}^{m} \frac{q^{(m-k)(1-2k)}(q^2;q^2)_m}{(q^2;q^2)_k(q^2;q^2)_{m-k}} \,z_1^k z_2^{m-k} =
q^m z_2^m + \frac{q^{1-m}(1-q^{2m})}{1-q^2} z_1 z_2^m + O(z_1^2)
$$
for the rightmost factor in (\ref{R2uv}) at $\varepsilon=0$.

Taking into account (\ref{qexpShort}) and $\overline{\mathbf{u}}^2 \cdot z_1 = 0$ we obtain
$$
\left(q^{u_2-v_1+1} \,\overline{\mathbf{u}}\, ; q^2\right)^{-1} \cdot z_1 =
z_1 - q^{u_2 -v_1-1}z_2\,,
$$
and finally after some trivial algebra we find the simple pole contribution
to $\mathrm{R}^2(v_1,u_2|v_2)\cdot z^m_1$ is equal
\begin{equation*}
\frac{(q^{2 v_1 - 2 u_2 + 4};q^2)q^{2-m}}{(q^{2\varepsilon};q^2)(1-q^2)}
\left[ q^{v_2-u_2-1}(1-q^{2m}) \cdot z_1 z_2^{m-1}-(1-q^{2 v_1 -2 u_2 +2+2m} )
\cdot z_2^m \right]+O(1)\,,
\end{equation*}
where the first term drops out at $m=0$.
Thus we are ready to calculate the restriction of the $\mathrm{R}$-operator
to the subspace (\ref{Fspace})
$$
z_1^m \to \frac{\left(q^{2(-u-\ell+\frac{1}{2})} ;q^2\right)}
               {\left(q^{2(u-\ell-\frac{1}{2})} ;q^2\right)} q^{-u-\ell-\frac{1}{2}}
               \left[ (q^{u+\ell-m+\frac{1}{2}}-q^{-u-\ell+m-\frac{1}{2}})
\cdot z_2^m - (q^m-q^{-m}) \cdot z_1 z_2^{m-1} \right],
$$
$$
z_1^m z_2 \to \frac{\left(q^{2(-u-\ell+\frac{1}{2})} ;q^2\right)}
               {\left(q^{2(u-\ell-\frac{1}{2})} ;q^2\right)} q^{-u-\ell-\frac{1}{2}}
               \left[ -(q^{m-2\ell}-q^{-m+2\ell}) \cdot z_2^{m+1} +
                       (q^{u+m-\ell+\frac{1}{2}}-q^{-u-m+\ell-\frac{1}{2}})
\cdot z_1 z_2^{m} \right].
$$
Taking into account permutation $\mathbb{R}_{12}= \mathrm{P}_{12}\mathrm{R}_{12}$
and choosing the basis in the space $\mathbb{C}^2$
as follows
$$
\mathbf{e}_1 = -z_2 \;\;;\;\; \mathbf{e}_2 = 1
$$
we finally conclude that the reduction of the $\mathrm{R}$-operator
coincides with $\mathrm{L}$-operator (\ref{Lq})
$$
\left.
\begin{array}{c}
\mathbb{R}_{12}\left(u\,|\,\ell,s=\frac{1}{2}\right)
\end{array}
\right|_{\mathbb{V}_{\ell}\otimes\mathbb{C}^2} =
\frac{\left(q^{2(-u-\ell+\frac{1}{2})} ;q^2\right)}
     {\left(q^{2(u-\ell-\frac{1}{2})} ;q^2\right)} q^{-u-\ell-\frac{1}{2}} \cdot
     \begin{array}{c}
     \mathrm{L}\left(u+\frac{1}{2}\,|\,\ell\right).
     \end{array}
$$
%%%%%%%%%%%%%%%%%%%%%%%%%%%%%%%%%%%%%%%%%%%%%%%%%%%%%%%%%%%%%%%%%%%%%%%%%%%%%%%%%%%%%%%%
%%%%%%%%%%%%%%%%%%%%%%%%%%%%%%%%%%%%%%%%%%%%%%%%%%%%%%%%%%%%%%%%%%%%%%%%%%%%%%%%%%%%%%%%
%%%%%%%%%%%%%%%%%%%%%%%%%%%%%%%%%%%%%%%%%%%%%%%%%%%%%%%%%%%%%%%%%%%%%%%%%%%%%%%%%%%%%%%%
\section{From the Coxeter relation to the explicit  $\mathrm{Q}$-operator
action}
\setcounter{equation}{0}
\label{altern}
%%%%%%%%%%%%%%%%%%%%%%%%%%%%%%%%%%%%%%%%%%%%%%%%%%%%%%%%%%%%%%%%%%%%%%%%%%%%%%%%%%%%%%%%
Here we present an alternative proof of the local relation (\ref{qR2GenFun}) which
produces an explicit formula of the Baxter $\mathrm{Q}$-operator (\ref{qQ}).
The present derivation uses  Coxeter relation (\ref{S3S2S3}) only.
At first taking into account the factorization of $\mathrm{R}^2$ (\ref{R2})
we rewrite (\ref{qR2GenFun})
as follows
\begin{equation} \label{R2gf}
\frac{(q^{2 z_1 \partial_{z_1} + 2 + 2b};q^2)}{(q^{2 z_1 \partial_{z_1}+2 + 2a + 2b};q^2)} \cdot
\frac{(q^{1-b}\frac{z_2}{z_1};q^2)}{(q^{1+2a+b}\frac{z_2}{z_1};q^2)}
\frac{(q^{2-a+b}\frac{z_1}{z_3};q^2)}{(q^{-a-b}\frac{z_1}{z_3};q^2)}
=
\frac{( q^{2+2b} , q^{1-b} \frac{z_2}{z_1} , q^{1-a} \frac{z_2}{z_3} , q^{2+a+b} \frac{z_1}{z_3} ; q^2)}
     {( q^{2+2a+2b} , q^{1+b} \frac{z_2}{z_1} , q^{1+a} \frac{z_2}{z_3} , q^{-a-b} \frac{z_1}{z_3}
;q^2)}\,.
\end{equation}
%%{\renewcommand{\arraystretch}{1.25}
%%\begin{equation} \label{R2gf}
%%\frac{(q^{2 z_1 \partial_{z_1} +2+ 2b};q^2)}{(q^{2 z_1 \partial_{z_1} +2+ 2a + 2b};q^2)} \cdot
%%\begin{array}{cccc}
%%( q^{1-b}\frac{z_2}{z_1} & , & q^{2-a+b}\frac{z_1}{z_3} & ;q^2) \\
%%\cline{1-4}
%%( q^{1+2a+b}\frac{z_2}{z_1} & , & q^{-a-b}\frac{z_1}{z_3} & ;q^2)
%%\end{array}
%%%\frac{(q^{1-b}\frac{z_2}{z_1};q^2)}{(q^{1+2a+b}\frac{z_2}{z_1};q^2)}
%%%\frac{(q^{2-a+b}\frac{z_1}{z_3};q^2)}{(q^{-a-b}\frac{z_1}{z_3};q^2)}
%%=
%%%\frac{( q^{2+2b} , q^{1-b} \frac{z_2}{z_1} , q^{1-a} \frac{z_2}{z_3} , q^{2+a+b} \frac{z_1}{z_3} ; q^2)}
%%%     {( q^{2+2a+2b} , q^{1+b} \frac{z_2}{z_1} , q^{1+a} \frac{z_2}{z_3} , q^{-a-b} \frac{z_1}{z_3} ;q^2)}
%%\begin{array}{cccccccc}
%%(q^{2+2b} & , & q^{1-b} \frac{z_2}{z_1} & , & q^{1-a} \frac{z_2}{z_3} & , & q^{2+a+b} \frac{z_1}{z_3} & ; q^2) \\
%%\cline{1-8}
%%(q^{2+2a+2b} & , & q^{1+b} \frac{z_2}{z_1} & , & q^{1+a} \frac{z_2}{z_3} & , & q^{-a-b} \frac{z_1}{z_3} & ; q^2)
%%\end{array}
%%\end{equation} }
To prove this we start with the Coxeter relation (\ref{S3S2S3}) and substitute the
explicit expression
for $\mathrm{S}^2$, $\mathrm{S}^3$ (\ref{qW}), (\ref{qS2})
%%Now we are going to show that the formula (\ref{R2gf})
%%can be deduced formally from the Coxeter relation (\ref{S3S2S3}):
\begin{equation} \label{s3s2s3=s2s3s2}
\frac{(\mathbf{v}q^{2b};q^2)}{(\mathbf{v} q^{2a+2b};q^2)}
\frac{(\mathbf{u}q^{-b};q^2)}{(\mathbf{u}q^{2a+b};q^2)}
\,q^{-(a+b)z_1 \partial_{z_1}}
\frac{(\mathbf{v};q^2)}{(\mathbf{v}q^{2b};q^2)}
=
\frac{(\mathbf{u}q^{-b};q^2)}{(\mathbf{u} q^{b};q^2)}
\frac{(\mathbf{v} ;q^2)}{(\mathbf{v} q^{2a+2b};q^2)}\, q^{-(a+b)z_1 \partial_{z_1}}
\frac{(\mathbf{u} q^{-a};q^2)}{(\mathbf{u} q^{a};q^2)}\,,
\end{equation}
where $\mathbf{u}$ and $\mathbf{v}$ form a Weyl pair:
$
\mathbf{u} = \frac{z_2}{z_1}q \;\;;\;\; \mathbf{v} = q^{2 z_1 \partial_{z_1}+2} \;\; ; \;\;
\mathbf{u} \, \mathbf{v} = q^2 \, \mathbf{v} \, \mathbf{u}
$.
Then we apply both sides of the identity (\ref{s3s2s3=s2s3s2}) to the
delta function in series expansion
$
\sum^{+\infty}_{k=-\infty} \left(\frac{z_1}{z_3}\right)^k = z_3 \delta(z_1-z_3)
$. The series represents actually the $\delta$ distribution on the unit
circle, i.e. with the restriction to the variables $|z_i| = 1$. Here we apply
the series expansion formally ignoring this restriction.

Thus in the left hand side we have
$$
q^{-(a+b)z_1 \partial_{z_1}}
\frac{(\mathbf{v};q^2)}{(\mathbf{v}q^{2b};q^2)} \cdot z_3 \delta(z_1-z_3)
= \sum_{k\geq 0} \frac{(q^{2+2k};q^2)}{(q^{2+2b+2k};q^2)} \, q^{-(a+b)k} \left(\frac{z_1}{z_3}\right)^k =
\frac{(q^2;q^2)}{(q^{2+2b};q^2)}
\frac{(\frac{z_1}{z_3}q^{2-a+b};q^2)}{(\frac{z_1}{z_3}q^{-a-b};q^2)}
$$
where $q$-binomial formula (\ref{qBinom}) is used on the last step.
Let us note that summation over all integer $k$ reduces
to summation over non-negative $k$ due to the factor $(q^{2+2k};q^2)$.
Similarly in the right hand side we have
%\begin{equation}
\begin{gather*}
\frac{(\mathbf{v} ;q^2)}{(\mathbf{v} q^{2a+2b};q^2)}\, q^{-(a+b)z_1 \partial_{z_1}}
\frac{(\mathbf{u} q^{-a};q^2)}{(\mathbf{u} q^{a};q^2)} \cdot z_3 \delta(z_1-z_3) = \\ =
\left.\frac{(\mathbf{u} q^{-a};q^2)}{(\mathbf{u} q^{a};q^2)}\right|_{z_1\to z_3}
\sum_{k\geq 0} \frac{(q^{2+2k};q^2)}{(q^{2+2a+2b+2k};q^2)} q^{-(a+b)k} \left(\frac{z_1}{z_3}\right)^k
=%\\=
\frac{(\frac{z_2}{z_3}q^{1-a};q^2)}{(\frac{z_2}{z_3}q^{1+a};q^2)}
\frac{(q^2;q^2)}{(q^{2+2a+2b};q^2)}
\frac{(\frac{z_1}{z_3}q^{2+a+b};q^2)}{(\frac{z_1}{z_3}q^{-a-b};q^2)}.
\end{gather*}
%\end{equation}
Comparing both sides of the equation we finally arrive at (\ref{R2gf}).
%%%%%%%%%%%%%%%%%%%%%%%%%%%%%%%%%%%%%%%%%%%%%%%%%%%%%%%%%%%%%%%%%%%%%%%%%%%%%%%%%%%%%%%%
%%%%%%%%%%%%%%%%%%%%%%%%%%%%%%%%%%%%%%%%%%%%%%%%%%%%%%%%%%%%%%%%%%%%%%%%%%%%%%%%%%%%%%%%
%%%%%%%%%%%%%%%%%%%%%%%%%%%%%%%%%%%%%%%%%%%%%%%%%%%%%%%%%%%%%%%%%%%%%%%%%%%%%%%%%%%%%%%%

\end{document}